\documentclass[aps,preprint,superscriptaddress]{revtex4}
\usepackage{color,bm,epsfig,graphics,amssymb,amsmath,subeqnarray,setspace,graphicx,amsthm,epstopdf,subfigure,color}

\begin{document}
\title{Locomotion by tangential deformation in a polymeric fluid}
\author{Lailai Zhu}
\affiliation{Linn\'e Flow Centre, KTH Mechanics, S-100 44 Stockholm, Sweden}
\author{Minh Do-Quang}
\affiliation{Linn\'e Flow Centre, KTH Mechanics, S-100 44 Stockholm, Sweden}
\author{Eric Lauga}
\email{elauga@ucsd.edu}
\affiliation{Department of Mechanical and Aerospace Engineering, University of California San Diego, 9500 Gilman Drive, La Jolla CA 92093-0411, USA.}
\author{Luca Brandt}
\email{luca@mech.kth.se}
\affiliation{Linn\'e Flow Centre, KTH Mechanics, S-100 44 Stockholm, Sweden}
\date{\today}

\begin{abstract}
In several biologically relevant situations,  cell locomotion  occurs in
polymeric 
fluids with Weissenberg { number} larger than one. Here we present results of
three-dimensional numerical simulations for the steady  locomotion of a
self-propelled body in a model polymeric (Giesekus) fluid at low Reynolds
number. Locomotion is driven by steady tangential  deformation at the surface of
the body (so-called squirming motion).  In the case of a spherical squirmer, we
show that the swimming velocity is systematically less than that in a Newtonian
fluid, with a minimum occurring for Weissenberg numbers of order one. The rate
of work done by the swimmer always goes up compared to that occurring in the
Newtonian solvent alone, but is always lower than the power necessary to swim in
a Newtonian fluid with the same viscosity. The swimming efficiency, defined as
the ratio between the rate of work necessary to pull the body at the swimming
speed in the same fluid  and the rate of work done by swimming, is found to
always be increased in a polymeric fluid. Further analysis  reveals that 
polymeric stresses break the Newtonian front-back symmetry in the flow profile
around the body. In particular,  a strong negative elastic wake is present
behind the swimmer,  which correlates with strong polymer stretching, and its
intensity increases with Weissenberg number and viscosity contrasts.  {The
velocity induced by the squirmer is found to decay in space faster than} in a
Newtonian flow, with a strong { dependence} on the polymer relaxation time and
viscosity. Our computational results are also extended to  prolate spheroidal
swimmers and smaller polymer stretching are obtained for slender shapes compared
to bluff swimmers. The swimmer  with an aspect ratio of two is found to be the
most hydrodynamically efficient.

\end{abstract}

\maketitle

\section{Introduction}

Small organisms  displaying the ability to move usually do so in the presence of a viscous fluid \cite{vogel96}. This is the case, in particular, for swimming cells such as bacteria, protozoa, or  spermatozoa, which exploit the viscous forces induced by the movement of appendages such as flagella or cilia in order to propel themselves in a fluid environment \cite{braybook, bergbook}.

The peculiar fluid mechanics properties at low Reynolds numbers, which is the regime in which motile cells live, dictate the manner in which they are able to swim \cite{purcell77}.  Classical work emphasized the relationship between the time-varying deformation of the cell bodies and appendages, and their swimming and transport kinematics, for a variety of cell families   \cite{lighthill76, brennen77, childress81}. More recent work has focused on nonlinear aspects  such as cell-cell interactions, and the  coupling between external mechanical forces and internal biophysical activity \cite{lp09}. For example, the role of hydrodynamic interactions in  collective modes of locomotion has been the focus of much  work  \cite{simha02, dombrowski04, cisneros07, sokolov07, saintillan_POF}. In addition to their relevance to biology, the physical principles of cell locomotion has allowed for the design of synthetic swimming devices on small scales \cite{arai01, arai01_2, dreyfus05, behkam06, behkam07}.

One topic of renewed interest concerns the locomotion of biological cells in complex (non-Newtonian) fluids. As a counterpart to large organisms known to  deal with non-Newtonian fluids, most notably  gastropods crawling on pedal mucus  \cite{trueman75, denny80a, denny80b}, in many instances  eukaryotic or prokaryotic cells move in fluids displaying time-dependent and nonlinear rheological properties \cite{bird76, birdvol1, birdvol2, tanner88, doi88, larson99}.
Examples include the progression of spermatozoa through the cervical mucus of mammals and along the mucus-covered fallopian tubes \cite{katz78, katz80, katz81, suarez92, suarez06, fauci06}, or the locomotion of bacteria through host mucus and tissues \cite{MontecuccoRappuoli2001,WolgemuthCharonGoldstein2006}. Bacteria in biofilms are also embedded in a viscoelastic matrix \cite{o'toole00, donlan02, costerton87, costerton95}.

In these  instances where locomotion occurs in a non-Newtonian fluid,  one can define a dimensionless  Weissenberg number for the flow, $We$, defined as the product of the fluid relaxation time scale with the typical shear rate in the flow \cite{birdvol1, birdvol2, tanner88}. In many cases, $We \gtrsim 1 $ \cite{lauga07, Fu08, Fu_shape, FuWolgemuthPowers2009, laugaEPL, teran_PRL}, indicating that elastic effects should play an important role in the  distribution of forces acting on cells.

A number of theoretical models have been proposed in the past to study small-scale locomotion in complex fluids. Linearized approach have used integral \cite{chaudhury79} or differential constitutive relationships \cite{fulford98}. Since cells relies on geometrical nonlinearities to  swim ---  a waving flagellum leads to locomotion at a speed scaling with the square of the wave amplitude   \cite{lp09} --- 
 nonlinearities in the constitutive modeling are essential, a result which has prompted renewed modeling interest.  Small-amplitude theories for swimming sheets
\cite{lauga07}, filaments \cite{Fu08, Fu_shape, FuWolgemuthPowers2009}, and arbitrary surface deformations \cite{laugaEPL} were recently obtained. Similarly, force generation arising from simple one-degree-of-freedom actuation modes were characterized  \cite{normand08, pak10}.

In practice, finite-size cells swim with large-amplitude motion, and in three dimensions. There is therefore fundamental interest in characterizing locomotion kinematics and energetics in  cases for which analytical treatment is not possible.  A recent numerical study in two dimensions addressed the locomotion of waving  sheets of large-amplitude, showing in particular that swimmers with non-constant wave amplitude could be more efficient and swim faster than their Newtonian counterparts  \cite{teran_PRL}. 

In this paper we take a further step in this direction. We present results of numerical simulations  for a steady squirmer free-swimming in  a model (Giesekus) polymeric fluid. Locomotion is achieved by steady tangential surface deformation of the cell, which displays no shape change. It is thus a model for locomotion by cells which swim using the propulsion generated by large arrays of short cilia \cite{blake74}, and is akin to the spherical envelope approach first proposed by Blake \cite{Blake1971a}. To the best of our knowledge, the results we present below are the first three-dimensional simulations for self-propelled motion in a complex fluid.

This paper is organized as follows. In Sec.~\ref{formulation} we present the modeling approach chosen in this paper, both for the swimmer and the complex fluid dynamics. 
In Sec.~\ref{numerics} we detail the numerical method used in our work, and the validation of the code. The main results are then presented in Sec.~\ref{spherical}, where we consider the case of a spherical squirmer, and present both integral properties of the locomotion (swimming speed, energetics, efficiency) as well as detailed flow characteristics (polymeric wake, flow streamlines,  and polymer stretching). A generalization to prolate swimmers is also offered in Sec.~\ref{prolate}. Our results are finally discussed in Sec.~\ref{discussion}.

\section{Problem formulation}
\label{formulation}

\subsection{Swimmer model}

In order to focus on the fundamental physics of locomotion in polymeric fluids, a  model microorganism is  used in this paper with several simplifying assumptions.  First, we mostly assume the microorganism to be spherical in  shape, as is the case for some unicellular ciliates such as {\it  Opalina}, or multicellular algae such as {\it  Volvox}  \cite{brennen77,Larson-Kirk}. Other organisms, such as {\it Paramecium} or {\it Cyanobacteria}  \cite{ratio2swim} have elongated shapes, and thus we also consider prolate ellipsoidal swimmers of varying aspect ratios in  Sec.~V. Second, swimmers are considered to be neutrally buoyant as their sedimentation velocity is much smaller than their swimming speed \cite{swimInter-Pedley}. Third, the small swimming speed and cell size make it reasonable to neglect inertial effects in the  flow, as commonly done in investigations of small-scale biological locomotion \cite{swimInter-Pedley,teran_PRL,lauga07,lp09}. Finally, Brownian effects are neglected, an assumption which is valid for all but the smallest bacteria. 

In this paper, the swimmers self-propel by generating tangential surface motion, as a model for the time-averaged ciliary propulsion by means of synchronized beating arrays of cilia \cite{blake74}. This forms the basis of the  so-called envelope model, as first introduced by Blake \cite{Blake1971a},  where the dynamics of the  ciliary tips are replaced by that of their continuous envelope. In this approximation,  an effective non-homogenous boundary condition is imposed at a fixed outer surface, which is  impermeable to the fluid. For the simulations presented in this paper, the surface velocity is assumed to be axisymmetric and time independent. That second assumption is justified if we are interested in the mean motion of a cell averaged over a fast beating period \cite{swimInter-Pedley}. This model microorganism is also referred to as ``squirmer'' in the literature. 

\begin{figure}[t]
   \centering
   \includegraphics[width=0.45 \textwidth]{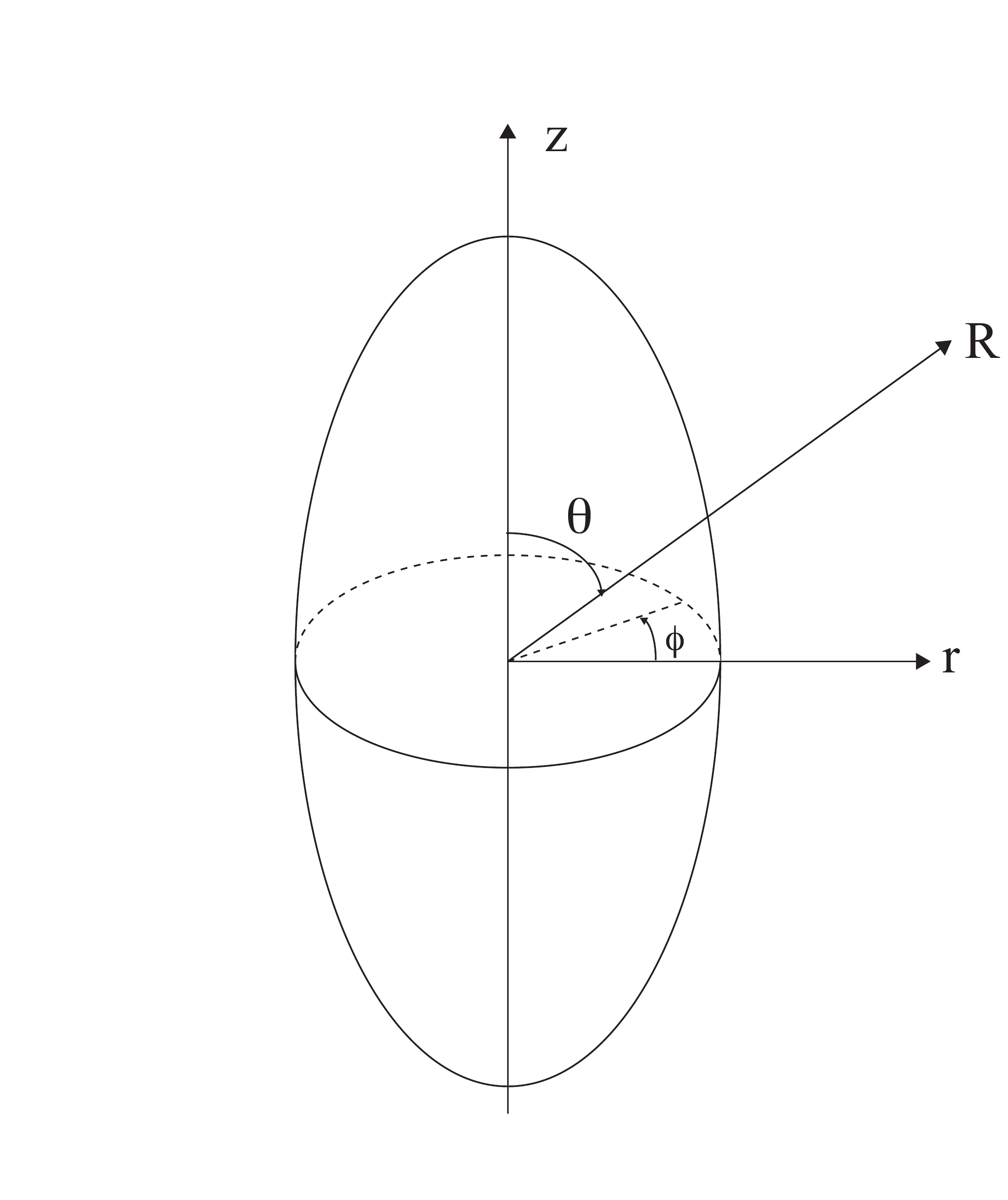} 
  \caption{Sketch of the prolate swimmer and the coordinate system used; ($r$, $z$) denote cylindrical coordinates and ($R$, $\theta$) the spherical coordinates with $\phi$ the azimuthal angle. The results presented below assume  axisymmetric flow.}
   \label{fig:coordisys}
\end{figure}

The surface velocity of a squirmer $\mathbf{u}_S$, the co-moving frame, is that considered Blake \cite{Blake1971a} with a concise formulation introduced in Ref.~\cite{PhysRevLett.100.088103} as
\begin{equation}
 \mathbf{u}_{\small{S}} ({\bf R})=\sum_{n\geq1} \frac{2}{n(n+1)}B_n
  P_{n}^{'}\left(\frac{\bf{e} \cdot \mathbf{R}}{R}\right)
  \left( \frac{\mathbf{e} \cdot \mathbf{R}}{R} \frac{\mathbf{R}}{R}-\mathbf{e} \right).
\end{equation}
Here, $\mathbf{e}$ is the orientation vector of the squirmer, $B_n$ is the $n$th mode of the surface squirming velocity \cite{Blake1971a}, $P_{n}$ is the $n$th Legendre polynomial, $\mathbf{R}$ is the position vector, and $R=|\mathbf{R}|$. (See Fig.~\ref{fig:coordisys} for a sketch of the notations)
In a Newtonian fluid, the swimming speed of the squirmer is $2B_{1}/3$ \cite{Blake1971a} and thus only dictated by the first mode. 
In previous studies, it is commonly assumed $B_{n}=0$ for $n>2$ \cite{PhysRevLett.100.088103,SimulationModelSwimmer}. Consequently, the tangential velocity on the sphere in the co-moving frame is expressed as $u_{\theta}(\theta)=B_{1}\sin\theta+(B_{2}/2 )\sin2\theta$, where $\theta=\arccos (\mathbf{e} \cdot \mathbf{r}/r )$ and 
an additional parameter, representing the ratio of the second to the first squirming mode, is introduced $\beta_{SW}$, i.e., $\beta_{SW}=B_{2}/B_{1}$. A squirmer with positive $\beta_{SW}$ is a ``puller'', and has its propeller located ahead of the cell body, while a squirmer with negative $\beta_{SW}$ is a ``pusher'', and has its propeller located behind the cell body in the swimming direction  \cite{PhysRevLett.100.088103,lp09}. In this paper we limited ourselves to the simple case $\beta_{SW}=0$, which gives the most energy-saving swimming gait { in a Newtonian fluid} \cite{SimulationModelSwimmer}. This convenient mathematical assumption allows us to numerically explore a large range of values of polymeric elasticity and viscosity. Thus, the surface velocity on our squirmer is $u_{\theta} (\theta)= B_{1} \sin{\theta}$, and has its maximum surface velocity located at the equator. For the simulations of the prolate organisms we assume the same boundary condition for the velocity component tangential to the surface of the ellipsoid. 

\subsection{Polymeric fluid dynamics}

For incompressible low-Reynolds number flow in a viscoelastic fluid,  the momentum and continuity equation are written as
\begin{eqnarray} \label{eq:mom}
 -\nabla p + \nabla \cdot \boldsymbol{\tau}&=&0,\\ 
 \nabla \cdot \mathbf{u}&=&0,\label{eq:con}
\end{eqnarray}
upon nondimensionalizing velocity with $B_{1}$, length with the diameter of the squirmer $D$, time with $D/B_{1}$, and pressure and stresses with $\mu B_{1} / D$, where $\mu$ is the solution viscosity.
Note that for numerical convenience we actually retain the partial time derivative in the momentum equation and present the final steady state results \cite{tranterm,viscoeReview}. 
Following classical modeling approaches \cite{birdvol1,birdvol2,bird76}, 
the deviatoric stress $\boldsymbol{\tau}$ can be splited into two components, the viscous solvent stress ($\boldsymbol{\tau}^{s}$) and the polymeric stress ($\boldsymbol{\tau}^{p}$); $\boldsymbol{\tau}^{s}$ is thus given by
\begin{equation}
 \boldsymbol{\tau}^{s}=\beta (\nabla \mathbf{u} + \nabla \mathbf{u}^{T}),
\end{equation}
where $\beta  < 1$ represents the ratio of the solvent viscosity, $\mu_{s}$, to the total zero shear rate viscosity,  $\mu$.
To complete the model, a transport equation for the polymeric stress $\boldsymbol{\tau}^{p}$ is required. Here we adopt the nonlinear Giesekus model \cite{giesekus}, which, in addition to shear-thinning material properties,  provides two important features, namely saturation of polymer elongation, and a non-negative entropy production during the time evolution of the polymers (see details in Ref.~\cite{Beris-Edwards,Dupret-Marchal,Souvaliotis-Beris}). Violation of these two properties may cause numerical difficulties and non-physical flow behavior. The nondimensionalized constitutive equation can be written as
\begin{equation}\label{eq:poly}
 \frac{\boldsymbol{\tau}^p}{We}+\overset{\triangledown}{\boldsymbol{\tau}^p} + \frac{\alpha}{1-\beta}(\boldsymbol{\tau}^p \cdot\boldsymbol{\tau}^p)=\frac{1-\beta}{We}(\nabla \mathbf{u} + \nabla \mathbf{u}^{T}),
\end{equation}
where upper-convected derivative, $ \overset{\triangledown}{\mathbf{A}}$,  defined for a tensor $\mathbf{A}$, is given by
\begin{equation}
 \overset{\triangledown}{\mathbf{A}}=\frac{\partial \mathbf{A}}{\partial t}+\mathbf{u} \cdot \nabla \mathbf{A}-\nabla \mathbf{u} ^T \cdot \mathbf{A}-\mathbf{A} \cdot \nabla \mathbf{u}.
\end{equation}
In the expression above, $We$ is the Weissenberg number, defined as $We={\lambda B_{1}}/{D}$ where $\lambda$ is the polymer relaxation time. The so-called mobility factor $\alpha$ is introduced in the nonlinear stress term representing an anisotropic hydrodynamic drag on the polymer molecules \cite{birdvol1}, and it limits the extensional viscosity of the fluid. From thermodynamics considerations, the mobility factor $\alpha$ must be in the $0-0.5$ range \cite{birdvol1,Larson}. We fix it to be $0.2$ in all our simulations.

\subsection{Swimming power and efficiency}

In the realm of low Reynolds number locomotion, where inertia can be neglected, the swimming speed is determined at each instant  as the speed at which the total force on the microorganism is zero. Given that swimming speed, it is of interest to compute the power required to move, and the efficiency of the motion. 
The power $P$ consumed by a swimming microorganism is defined as \cite{PRLswim}
\begin{equation}
 P=-\int\!\!\!\int_{S}\mathbf{n} \cdot \boldsymbol{\sigma} \cdot \mathbf{u}\, {\rm d} S,
\end{equation}
where $\mathbf{n}$ is the unit normal outward the swimmer surface $S$, $\boldsymbol{\sigma}$ is the stress tensor, $\boldsymbol{\sigma}=-p\mathbf{I}+ \boldsymbol{\tau}^{s}+\boldsymbol{\tau}^{p}$, and $\mathbf{u}$ is the velocity of the fluid in the laboratory reference frame. The { swimming efficiency 
\begin{equation}\label{def_eta}
\eta= \frac{F_p \, U}{P}
\end{equation}
is then defined as the ratio between the work rate $F_p \, U $ necessary to pull the swimmer body at the swimming speed in the same fluid (same $We$ and $\beta$) without active boundary motion and the swimming power $P$ defined above \cite{lp09}.  
The force acting on the pulled object is given by
\begin{equation}
 F_p= - \int\!\!\!\int_{S}\mathbf{n} \cdot \boldsymbol{\sigma} \, {\rm d} S.
\end{equation}
}

\section{Numerical method}
\label{numerics}

The finite-element code Femlego, developed at { the Royal Institute of Technology (KTH), Stockholm} \cite{femlego} is used in our simulations. Femlego has provided a variety of successful simulations in the area of microfluids \cite{femwet} and multiphase flow \cite{femsplash}. For the incompressible isothermal Navier-Stokes equations, a projection method introduced by Guermond and Quartapelle \cite{proFem} is used  to solve the conservation equation for mass and momentum, Eqs. (\ref{eq:mom}) and (\ref{eq:con}), with a Galerkin discretization. However, Galerkin discretization is not the optimal choice for the consititutive equation,  Eq.~(\ref{eq:poly}), owing to the increasing importance of the convective term  with increasing Weissenberg number  \cite{MITnonElasticBound}. Serving as a remedy, we follow Marchal \cite{viscoeReview} and Frank  \cite{supgCrochet} and adopt the streamline-upwind/Petrov- Galerkin (SUPG) method for the convective term in the constitutive equation.
The weak form of the consitutive equation is therefore written as

\begin{equation}
 \left\{\mathbf{S}+\frac{h}{U}\mathbf{u} \cdot \nabla \mathbf{S},\mathbf{\tau^p}+\lambda (\overset{\triangledown}{\mathbf{\tau^p}}+\frac{\alpha}{\mu_p}(\mathbf{\tau^p} \cdot \mathbf{\tau^p}))-\mu_p(\nabla \mathbf{u}+\nabla \mathbf{u}^T )\right\}=0
\end{equation}
where $\mathbf{S}$ denotes the weighting function for $\tau_p$, $h$ is a characteristic length-scale of the element and $U$ is the magnitude of the local characteristic velocity. In our case, we choose the norm of $\mathbf{u}$ as the value of $U$. 

The fact that the flow is axisymmetric is exploited in our simulations, degenerating the computational domain to a half circle representing the squirmer bounded by a rectangular box. Inflow boundary is placed 10 diameters away from the object with prescribed velocity and zero polymeric stress (equilibrium status). The outflow boundary is 30 diameters downstream of the object, with zero pressure specified as the flow is fully developed. The centerline is treated as an axisymmetric boundary condition. Tangential velocity is imposed on the surface of the squirmer to realize the prescribed swimming gait, as discussed above. Neumann boundary conditions are set for the remaining variables. 

Spatial discretization is performed with piecewise linear functions for the whole set of equations. Use of triangular elements in our simulations enabled sufficient grid refinement to better capture the unique flow structure in the polymeric flow, such as  elastic boundary layers \cite{MITnonElasticBound} and elastic wake \cite{Elasticwake}. The number of elements typically used in our simulations is of about 90,000 with necessary grid refinement up to 150,000 elements to avoid numerical instabilities at higher Weissenberg number. Mesh independence of the results has been tested for the most difficult cases, and a relative error below $1\%$ has been observed for both swimming speed and power.

The computation of the swimming speed is based on the fact there is no net force or torque on self-propelled swimming microorganisms. Therefore, we performed simulations of the same squirmer with three different free-stream conditions (the simulations are performed in the comoving reference frame) and computed the hydrodynamic forces for the three cases. By interpolation, we are thus able to estimate the swimming speed as the free-stream velocity for which the total force on the body is zero. For all the cases, simulations are then performed at the estimated swimming speed in order to verify that the force were below a given tolerance and to provide a more accurate evaluation of the swimming power.

\section{Locomotion of a spherical squirmer}
\label{spherical}
\subsection{Integral quantities}

\begin{figure}[t]
   \centering
   \includegraphics[width=0.48 \textwidth]{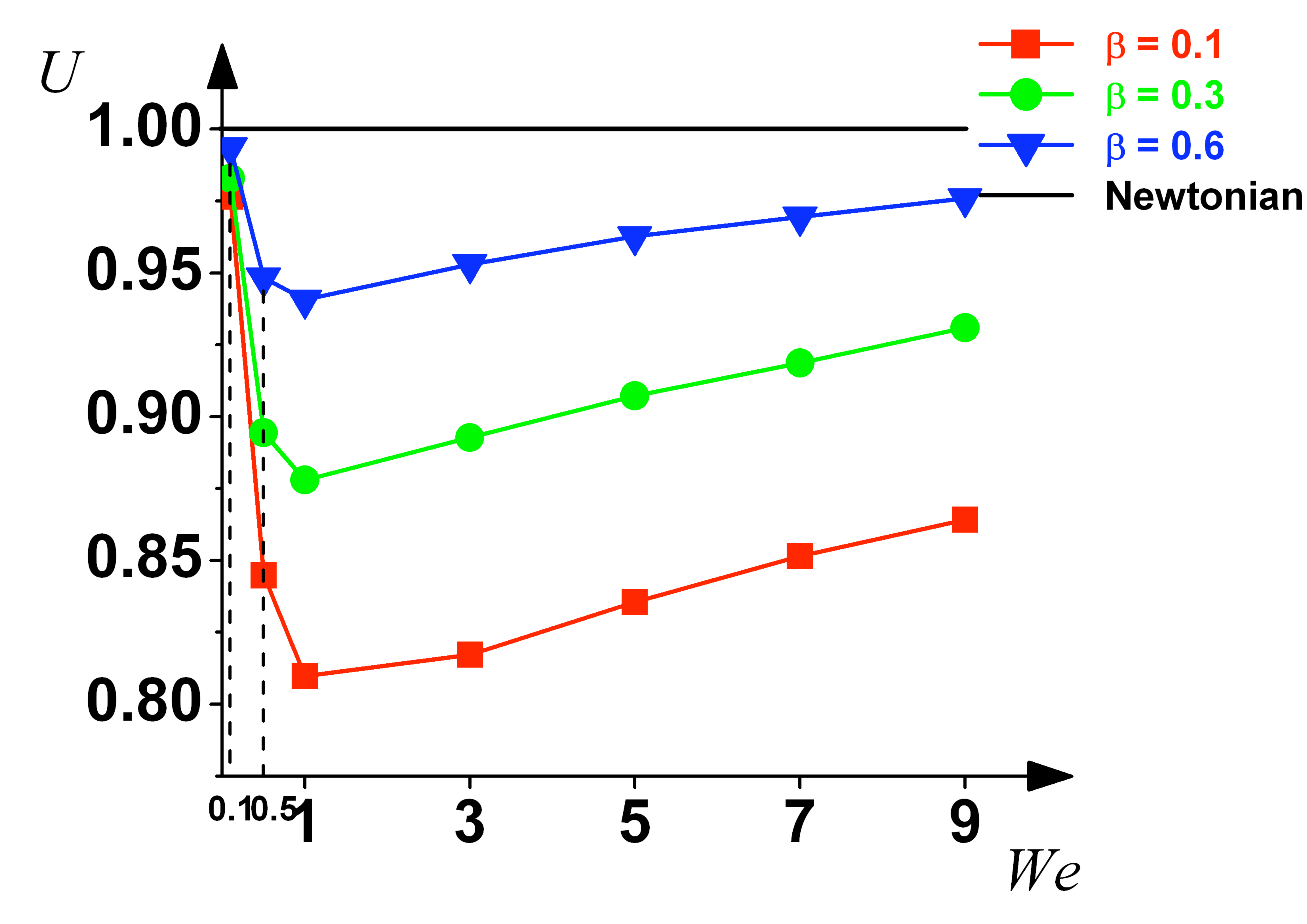} 
  \caption{(Color online) Swimming speed  $U$ in the polymeric fluid divided by that of the Newtonian swimmer, versus Weissenberg  number, $We$, for three values of the viscosity ratio, $\beta$: 0.1 (red squares), 0.3 (green circles), 0.6 (blue triangles). }
   \label{fig:speed}
\end{figure}

We first present our results on integral quantities of the swimming motion, namely the swimming speed, work done, and the swimmer efficiency. 
In the following the different quantities will be made non dimensional with the diameter of the spherical squirmer and its swimming velocity and power in the Newtonian fluid. Simulations are performed with different values of the Weissenberg  number, $We$, and for three values of the viscosity ratio, $\beta$ (0.1, 0.3, and  0.6). The swimming speed in the polymeric fluid divided by that of the Newtonian swimmer is displayed in Fig.~\ref{fig:speed} as a function of $We$. We see that the swimming speed of the squirmer decreases for low Weissenberg  numbers, reaches its minimum value near $We=1$, and then slowly recovers with increasing polymeric elasticity (or $We$). The largest decrease in swimming speed is observed for the lowest value of $\beta$ considered, i.e.~for the largest polymer viscosity under investigation. It is interesting to note that the minimum speed is always obtained when the polymer relaxation time is approximately equal to the time it takes for the swimmer to swim its own length.

\begin{figure}[t]
   \centering
   \includegraphics[width=0.48 \textwidth]{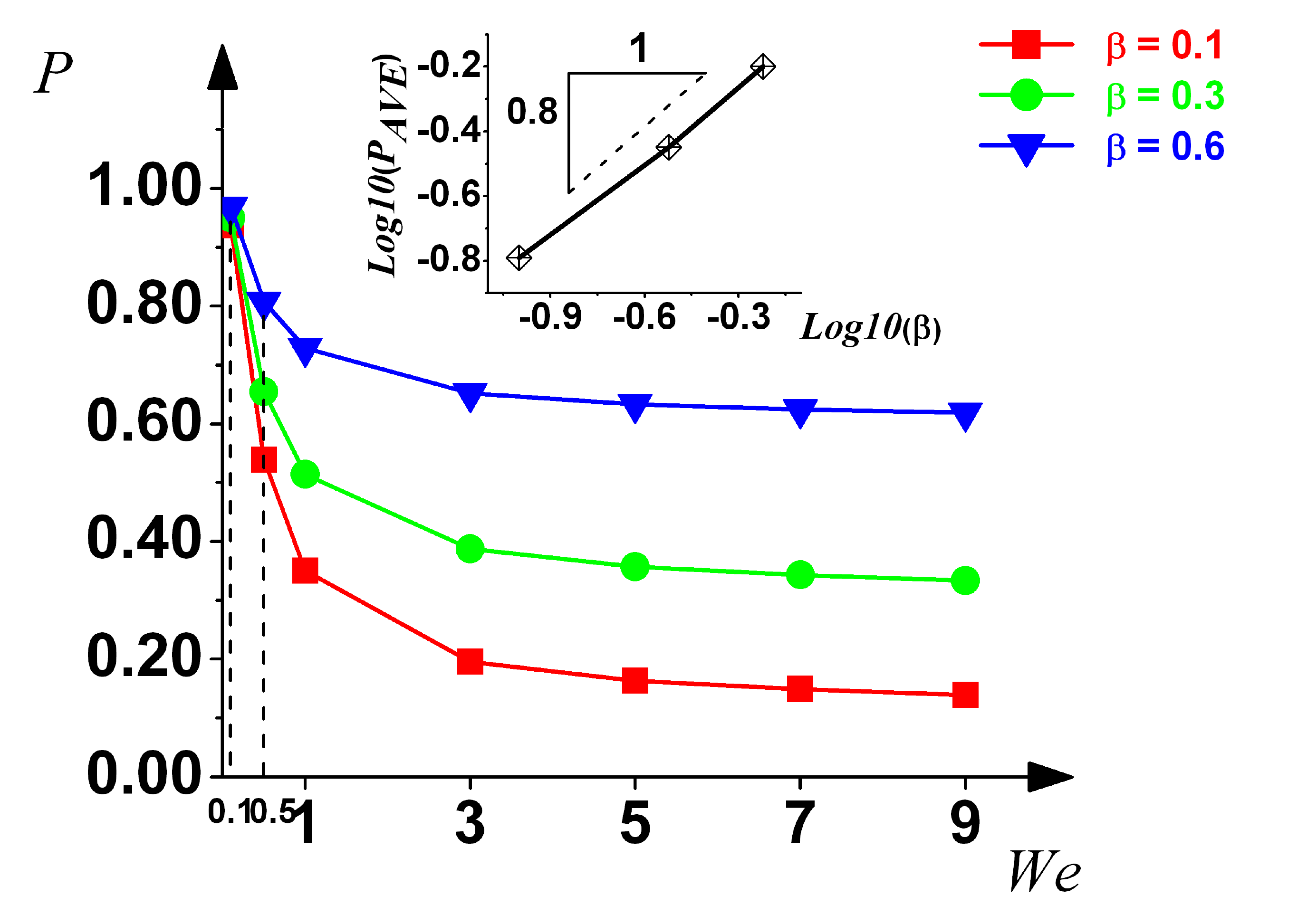}
  \caption{ (Color online) Swimming power $P$ divided by that in the Newtonian fluid with identical total viscosity, versus Weissenberg  number, $We$, for three values of the viscosity ratio, $\beta$: 0.1 (red squares), 0.3 (green circles), 0.6 (blue triangles). }
   \label{fig:power}
\end{figure}

The rate of work done by the swimmer divided by that in the Newtonian fluid with identical total viscosity  is displayed in Fig.~\ref{fig:power} as a function of $We$. The power needed to swim decreases for all cases considered and seems to approach a constant asymptotic value at large Weissenberg  number. It is important to note that, if scaled with the solvent viscosity, the actual value of the work performed by the microorganism is increasing when  decreasing $\beta$. However, the work done is significantly less that that of a swimmer in a Newtonian fluids with the same viscosity, similarly to what was observed for the swimming sheet \cite{lauga07}. This relative power saving in a polymeric fluid increases with  Weissenberg numbers in all cases. Note that the fact that the swimming speed and power approach the Newtonian values as $We\to0$ contributes to  an  {a posteriori} validation of our code.

The power expended by the squirmers is approaching a final value in the high-$We$ limit  independent of the fluid elasticity and increasing with the total viscosity. As will be discussed below, for long relaxation times, the stretching/relaxation of polymers in the wake of the body takes place further away from the organism. This may explain why the results become independent of the Weissenberg number: the elastic wake moves far enough not to affect the stress distribution close to the surface. At the same time, the work performed against the fluid is larger for larger viscosity and therefore the value of the power needed to swim when $We \to \infty$ increases when $\beta$ decreases. 
The swimming power is seen numerically to scale with $\beta$ approximately as $P \sim \beta^{-0.8}$.

\begin{figure}[t]
   \centering
   \includegraphics[width=0.48 \textwidth]{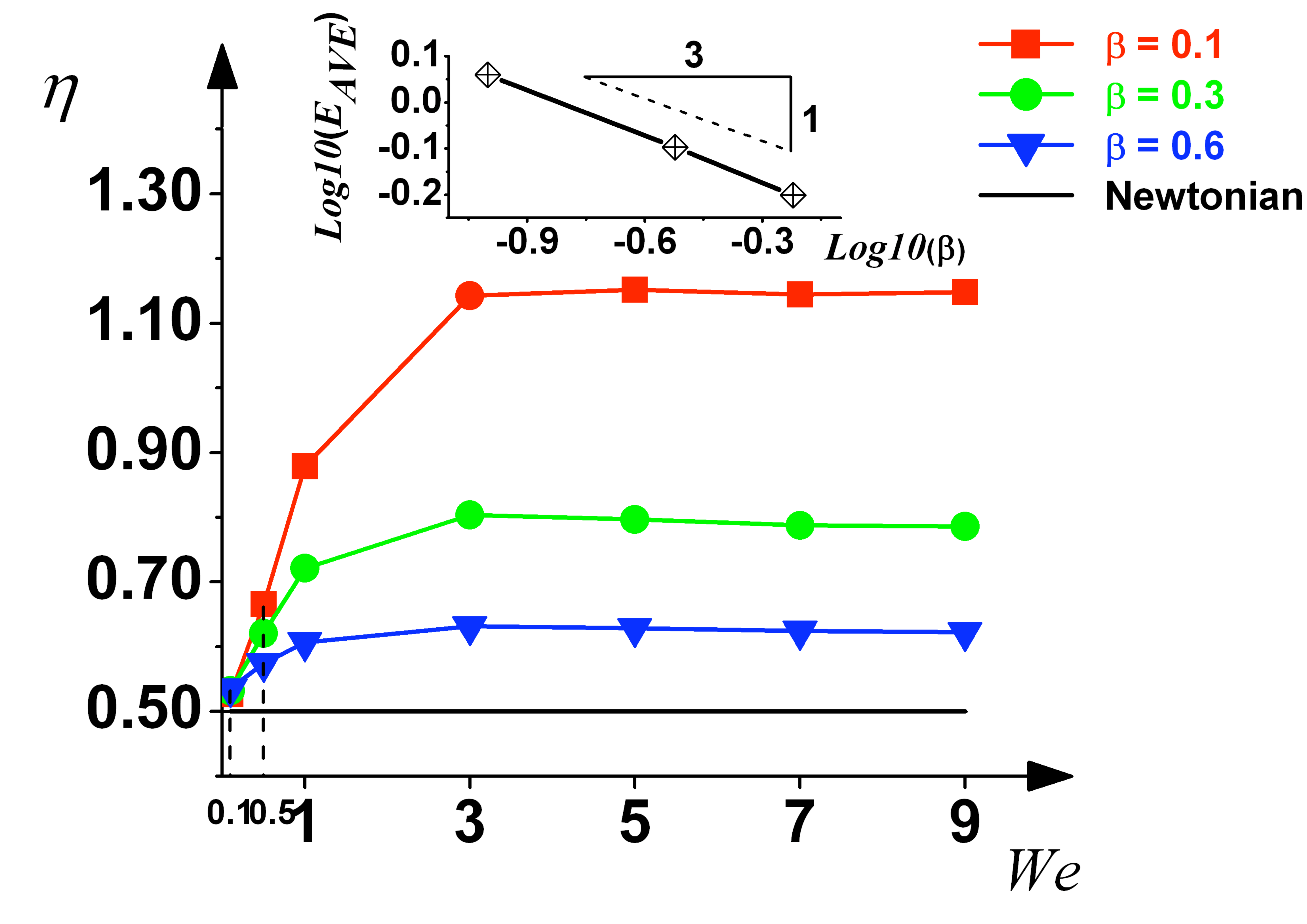}
   \caption{
   (Color online) Swimming efficiency, $\eta$, in a polymeric fluid: ratio between the power needed to pull the spherical body  at the velocity equal to  its swimming speed and  the power required to swim in the same fluid. Efficiency is displayed as a function of $We$ for three values of $\beta$.
   Inset: Value of the efficiency at large $We$ as a function of $\beta$ (log-log plot); the line is a guide for the eye showing a 1/3 power law.
   }
   \label{fig:eff}
\end{figure}

The swimming efficiency, $\eta$, is shown in Fig.~\ref{fig:eff} as a function of $We$. The efficiency is defined here as the ratio between the power needed to pull the spherical body at the  swimming velocity of the squirmer and the power required to swim in the same fluid. The efficiency is seen  to always be larger in the viscoelastic fluid than in a Newtonian fluid, which is one of the main results of our work. This is in agreement with the findings of Teran et al.  \cite{teran_PRL} who simulated a two-dimensional swimming sheet finite length in an Oldroyd B-fluid, as well as the results by Leshansky \cite{leshansky09} who considered the locomotion of a squirmer in a suspension of rigid spheres.  The efficiency is seen to remain essentially  constant beyond  $We\gtrsim3$. By considering the averaged values of the efficiency in the large-$We$ limit, the relation between the viscosity ratio and the asymptotic efficiency is examined. As shown by the inset in Fig.~\ref{fig:eff}, there seems to be a power-law relationship with exponent close to 1/3, $\eta \sim \beta^{-1/3}$. 
{ Using the definition of efficiency given in Eq.(\ref{def_eta}) and $P \sim \beta^{-0.8}$, we conclude that the relative decrease in power with viscosity ratio observed at large $We$ is faster for swimming micro-organisms than for pulled bodies.}

\subsection{Flow visualization}

\begin{figure}[t]
   \centering
   \includegraphics[width=0.4 \textwidth]{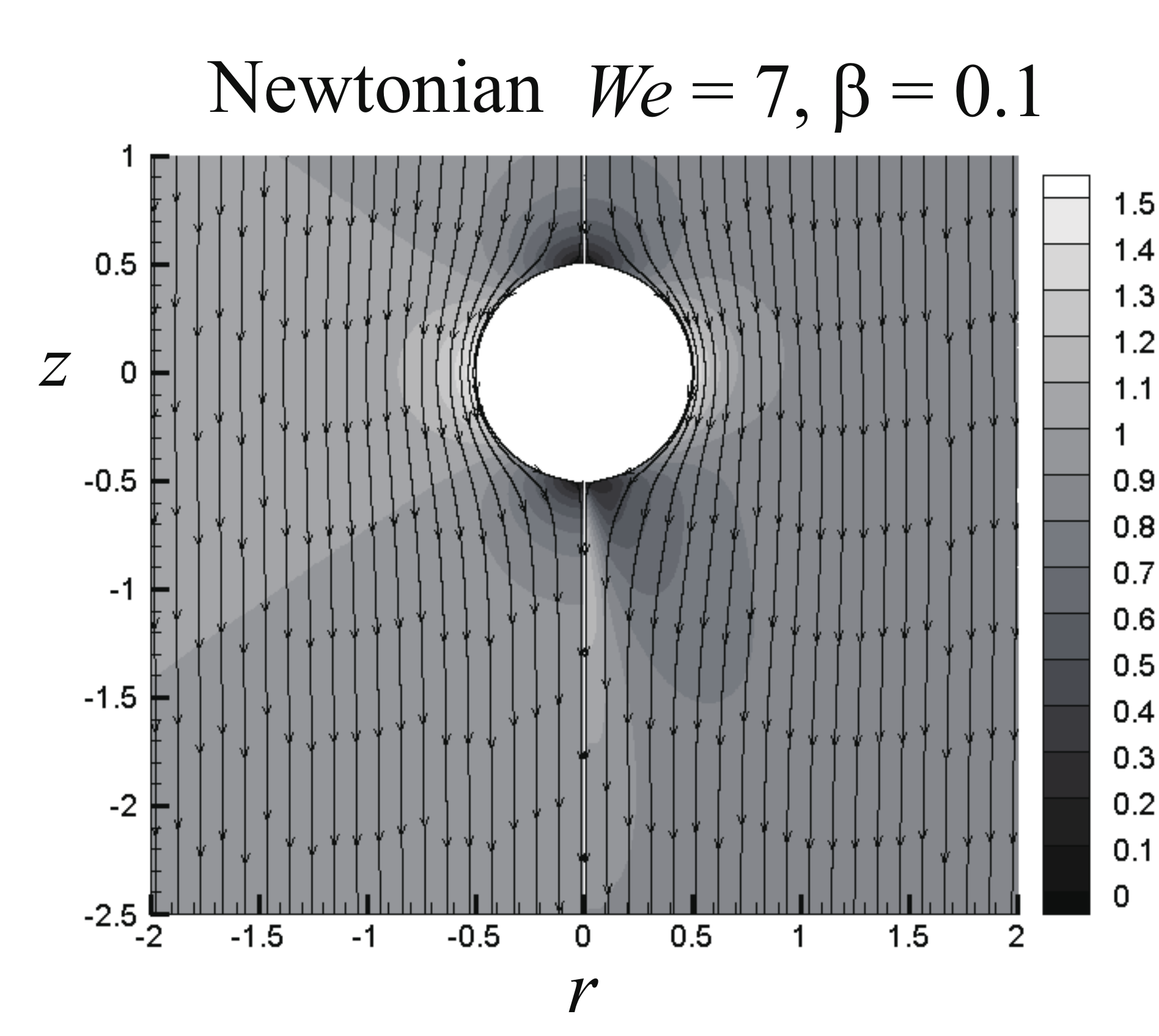}
\caption{Flow streamlines and  map of velocity magnitudes. Comparison between the Newtonian case (left) and polymeric case with $We=7$ and $\beta=0.1$ (right).}
   \label{fig:flow_newt}
\end{figure}

\subsubsection{Comparison with Newtonian swimming: Elastic wake}

In this section we consider the detailed flow (solvent plus polymer) around the swimming microorganism. We start by showing in Fig.~\ref{fig:flow_newt} the difference in velocity field between a Newtonian and a non-Newtonian squirmer. The figure depicts the Newtonian case (left-hand side) and  the polymeric flow with $We=7$ and $\beta=0.1$ (right-hand side), both in the co-moving frame. The streamlines are shown close to the body and the background  map indicates the velocity magnitude.

We first note the similarity in the shape of the  streamlines. The only noticeable difference is a slight  upstream shift upstream with increasing $We$ of the streamlines behind the cylinder. In the case of translation of a sphere, a similar observation has been attributed to the  shear-thinning characteristics of the viscosity, see among others Ref.~\cite{harlen02}.

The first important difference between Newtonian and polymeric swimming  is the magnitude of the fluid velocity. The flow approaching the swimmer is hardly changed by the presence of polymers, while important quantitative differences exist on the side and the front of the body. The velocity induced by the swimming gait, maximum in the equatorial plane $z=0$, decays faster in the viscoelastic fluid where a thinner boundary layer is observed. This fast  decay was  identified by Leshansky \cite{leshansky09} as a possible cause of larger efficiency for locomotion in a non-Newtonian media.

The second notable difference is  the presence of a so-called  negative elastic wake downstream of the object. This is clearly visible about one diameter behind the sphere, and it extends to about six diameters downstream for the longer polymer relaxation times. The front-back flow symmetry of a Newtonian swimmer is thus broken in a viscoelastic fluid. 

\begin{figure}[t]
   \centering
   \includegraphics[width=0.4 \textwidth]{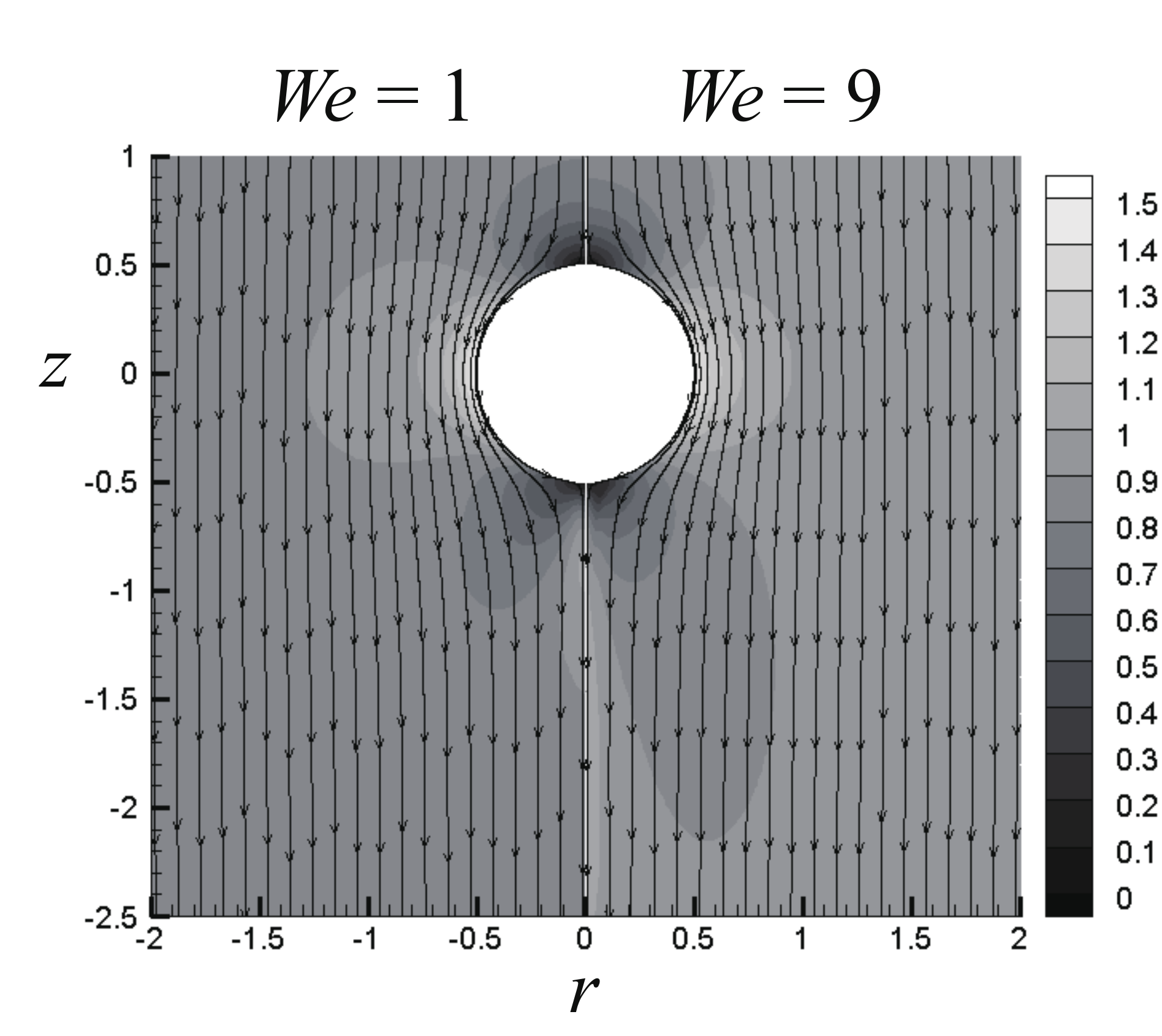}
\caption{Flow streamlines and  map of velocity magnitudes.Comparison between two polymeric fluids with same viscosity ($\beta=0.3$) with $We=1$ (left) and $We=9$ (right).}
   \label{fig:flow_We}
\end{figure}

The negative wake was studied for spheres sedimenting in polymeric flows \cite{hassager79,bisgaard83}. It  appears as a velocity overshoot behind the body in the co-moving reference frame and as a negative velocity in the laboratory frame, and is related to the relative magnitude of the normal and shear stress and their spatial gradients.  Stresses generated in the extensional flow at the rear of the squirmer drive the flow towards the body and produce a region of slower decay, the so-called extended wake. In contrast,  the force induced by the downstream relaxation of shear stresses generated near the side of the body  gives rise to flow directed away from the swimmer, and causes a negative wake \cite{harlen02}. Away from the axis of symmetry, the principal direction of extension is no longer aligned with the axis, which produces stresses directed away from the body. The polymers away from the axis have memory of the shear flow experienced near the side of the sphere and for large relaxation times the stresses built up in this region are still relevant as fluid particles are advected downstream.

\begin{figure}[t]
   \centering
   \includegraphics[width=0.45 \textwidth]{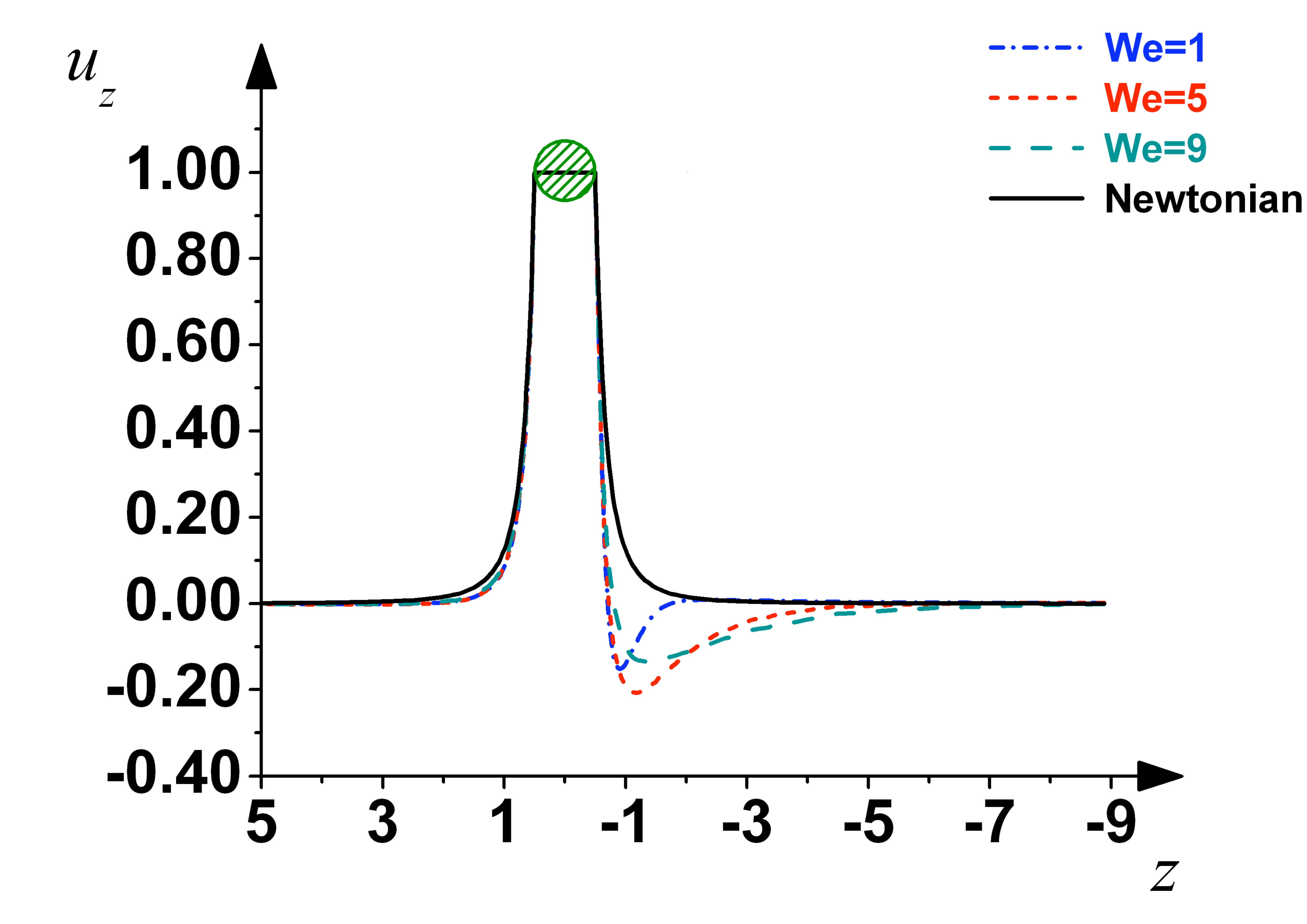}
   \caption{(Color online) Axial flow in front and behind the swimmer  nondimensionalized by the swimming speed of the organism for different values of the Weissenberg number and $\beta=0.3$.}
   \label{fig:wake_a}
\end{figure}

\begin{figure}[t]
   \centering
   \includegraphics[width=0.45 \textwidth]{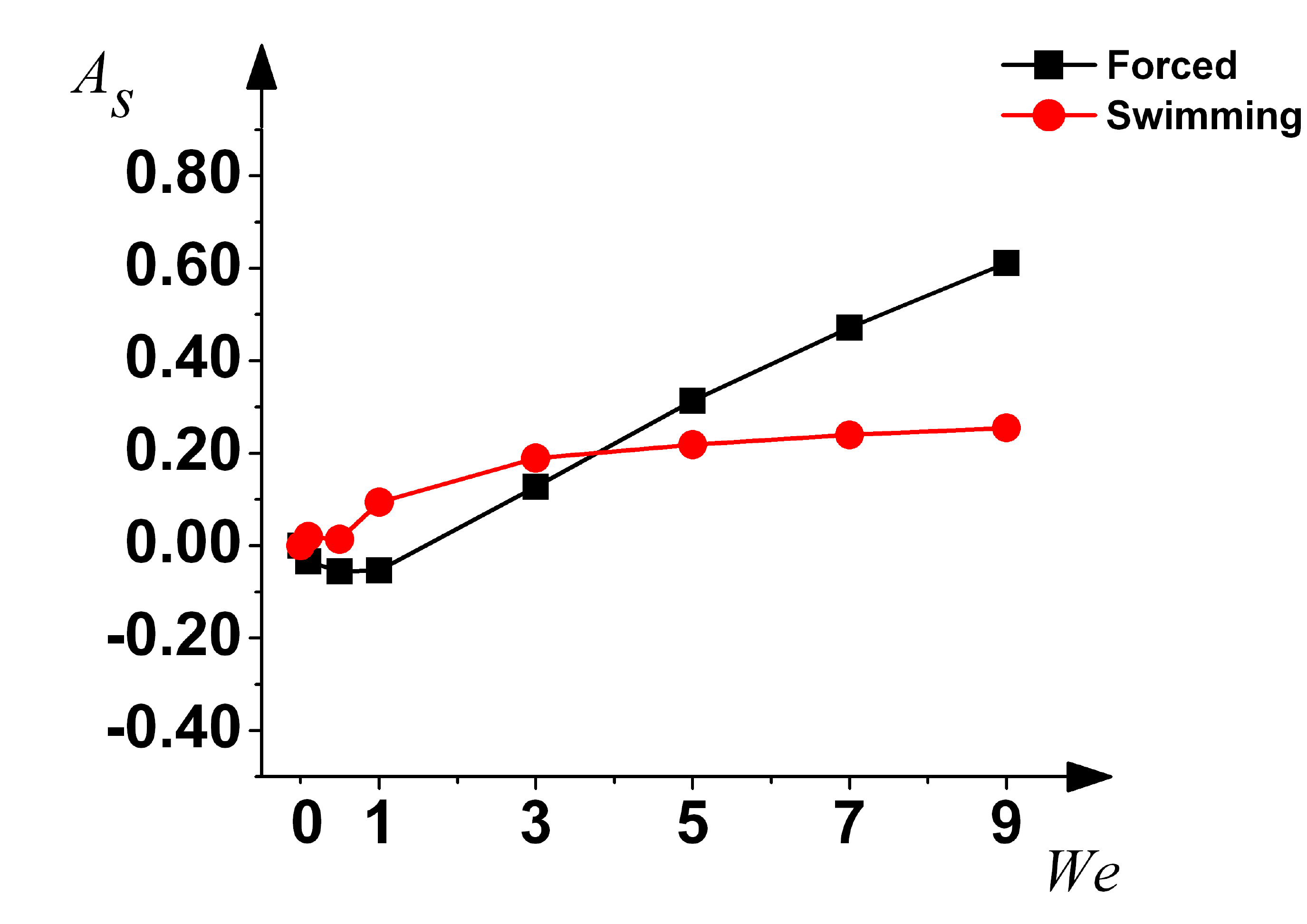}
   \caption{(Color online) Asymmetry measure (see text) as a function of the Weissenberg  number for swimming motion (red circles, solid line) and 
   forced motion at same speed with $\beta=0.3$ (black squares, solid line).}
   \label{fig:wake_b}
\end{figure}

A wake number is defined in Ref.~\cite{harlen02} to show that both limited polymer extension and large relaxation times contribute to the formation of a negative wake. The extensional viscosity plays an important role in the generation of the elastic wake, and a shear-thinning first normal stress coefficient enhances the velocity overshoot. Negative wakes are not expected in dilute solutions, such as those modeled  e.g. by the Oldroyd-B constitutive equation at moderate Weissenberg numbers. Fitting of experimental data of semi-concentrated solutions using Giesekus and Phan-Thien-Tanner models allow one to reproduce negative wakes in  numerical simulations \cite{arigo98}. Our results are thus relevant to locomotion in concentrated polymer solutions, and to relatively high values of the polymer relaxation times.

In Fig.~\ref{fig:flow_We} we further show  a comparison of the flow field for two different values of the Weissenberg number, $We=$ 1 and 9, at fixed viscosity ratio $\beta$. The elastic wake is evident for the largest $We$ considered while it appears not yet formed for $We=1$, as expected since $We$ is the ratio between the characteristic time scales for polymer relaxation and  advection.

\begin{figure}[t]
   \centering
   \includegraphics[width=0.4 \textwidth]{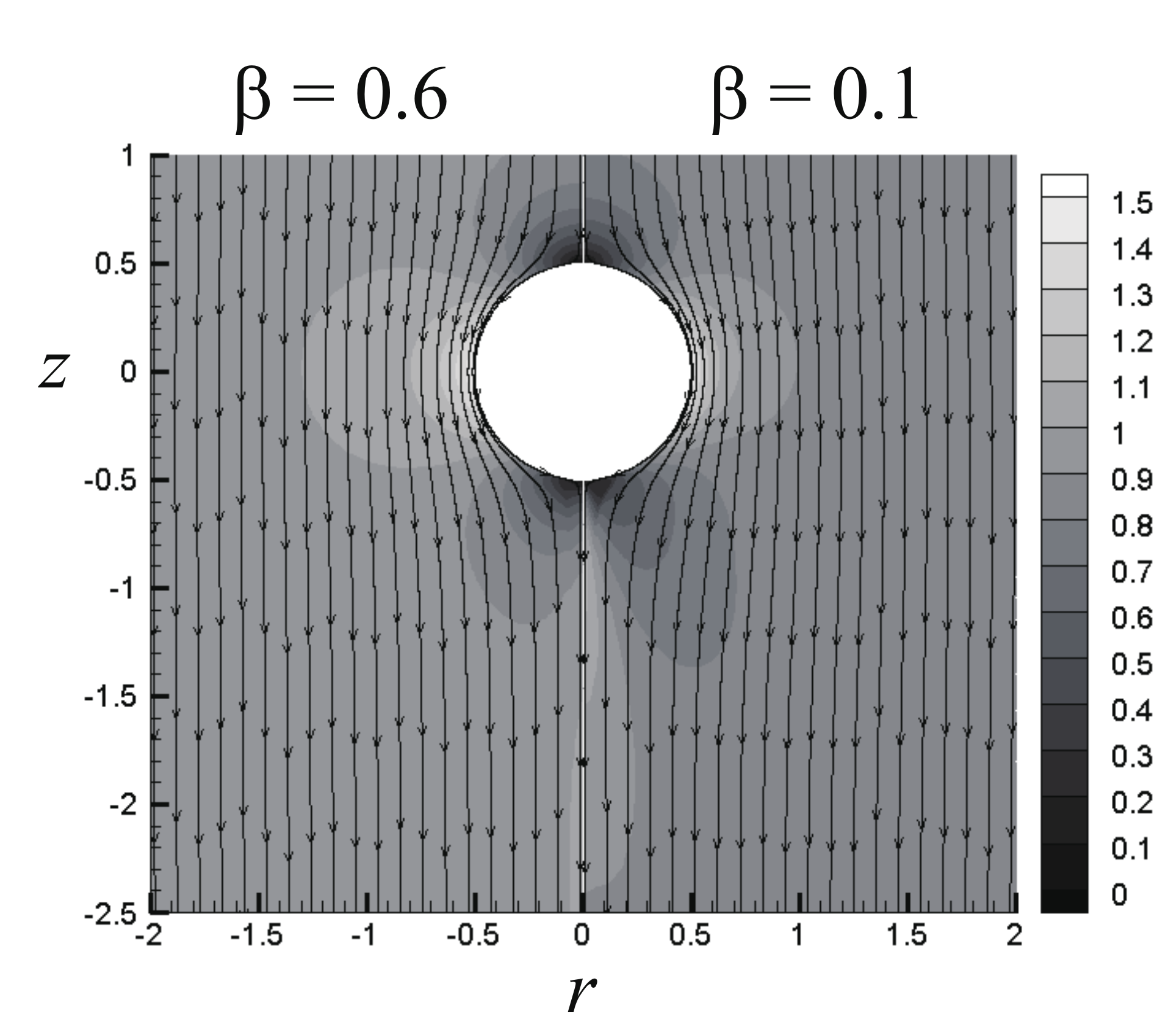}
   \caption{Influence of polymer viscosity on flow streamlines and velocity magnitudes for $We=7$. Left:   $\beta=0.6$. Right:  $\beta=0.1$.}
   \label{fig:flowb}
\end{figure}

To further investigate the occurrence of an elastic wake, we show the axial profile of the axial velocity along the symmetry axis $r=0$ in Fig.~\ref{fig:wake_a} divided by the swimming speed. 
The symmetric Newtonian case is also reported for comparison. For a sedimenting sphere, the extent of the negative velocity just downstream of the wake increases with $We$ (see Ref.~\cite{harlen02} and references therein). { In the polymeric fluid the largest negative value is found at $We \approx 5$.  This seems to result from two competing effects: 
When the Weissenberg number increases, the extent of the region of negative velocity increases as for the pulled object. However, the negative peak velocity decreases since the shear stress responsible for its formation is acting further downstream in the region of larger fluid velocity (seen in the co-moving frame)}. 

To quantify this effect, we consider as measure of the fore-aft asymmetry the difference in the axial velocity upstream and downstream of the object
\begin{equation}
\mathcal{A}_s=\int_{D/2}^{\infty} [ u_z(r) - u_z(-r) ] \, dz.
\end{equation}
 The variation of the asymmetry, $\mathcal{A}_s$, with the  Weissenberg number is shown in Fig.~\ref{fig:wake_b} at $\beta=0.3$ for both the swimmer and the forced motion of the spherical body  in the same polymeric solution. For the swimmer, the asymmetry, which is zero in the Newtonian limit, 
 always increases and reaches a constant value when $We\gtrsim5$. 
 As shown in Fig.~\ref{fig:wake_a}, this is due to a compensation between the elongation of the wake and the negative peak just behind the swimmer.  In the case of the forced motion of a sphere,  $\mathcal{A}_s$ is first negative for lower values of $We$ and then increases monotonically. This is explained by the fact that the decrease of the velocity in front of the object is faster for forced motion, and is observed already for the lowest Weissenberg number considered while the negative wake, leading to positive values of $\mathcal{A}_s$, is formed in this case only when $We\gtrsim3$.

Finally, the influence of polymer viscosity on flow streamlines and velocity magnitudes  is illustrated in Fig.~\ref{fig:flowb} for $We=7$.  
For increased flow viscosity,  the effect of swimming actuation on the side of the object is felt over shorter distances. Similarly, the  negative elastic wake is found about one diameter downstream of the swimmer at $\beta=0.1$, while it is further downstream, at $r/D \approx 2.5 $, for $\beta=0.6$.

\subsubsection{Spatial decay}

\begin{figure}[t]
   \centering
   \includegraphics[width=0.45 \textwidth]{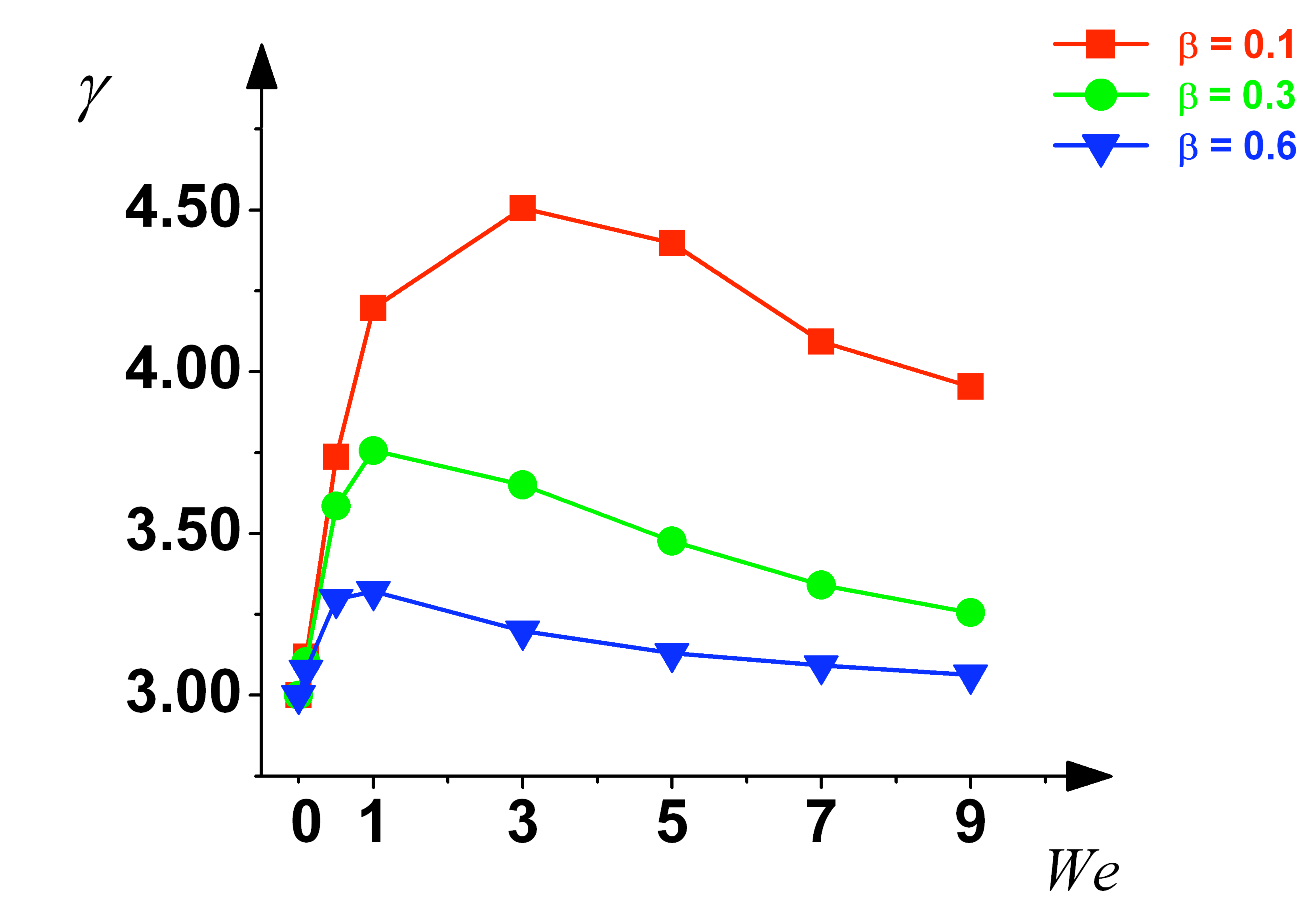}
   \caption{(Color online)  Power law exponent $\gamma$ for decay of the axial  velocity along the $z=0$ plane ($u \sim r^{-\gamma}$, see text), as a function of the Weissenberg  number.}
   \label{fig:decay}
\end{figure}

In order to quantify the spatial signature of the velocity perturbation introduced by the swimmer in directions other than that of the wake, we now consider the decay of the axial velocity in the radial direction along the equatorial plane, i.e. $u_z(r,z=0)$. For a Newtonian squirmer with $\beta_{SW}=0$ the velocity  decays as $\sim1/r^{3}$, whereas the decay is only $\sim1/r^2$ for pusher and puller-type cells ($\beta_{SW}\neq0$). We numerically estimate the radial decay of the velocity  for locomotion in a  polymeric fluid by fitting a power law from about $r \approx D$ to the end of the computational domain. The values of the exponent $\gamma$ obtained with this procedure are reported in Fig.~\ref{fig:decay}  as a function of the Weissenberg number. The flow, which decays as $\sim1/r^{3}$ in the Newtonian case, always decays faster in the polymeric case. 
 We observe that the variation of the decay rate with $We$ is not monotonic, and that for the two largest values of the viscosity ratio ($\beta=0.6, 0.3$) a maximum is reached near $We=1$, which coincides with the occurrence of the minimum swimming speed. Finally, the decay rate increases with increased  viscosity contrast between the polymer and the solvent (decrease of  $\beta$).
{ In agreement with Ref.~\cite{leshansky09} we find therefore that a more rapid decay leads to larger efficiency.}

\subsubsection{Polymer stretching}

\begin{figure}[t]
   \centering
   \includegraphics[width=0.48 \textwidth]{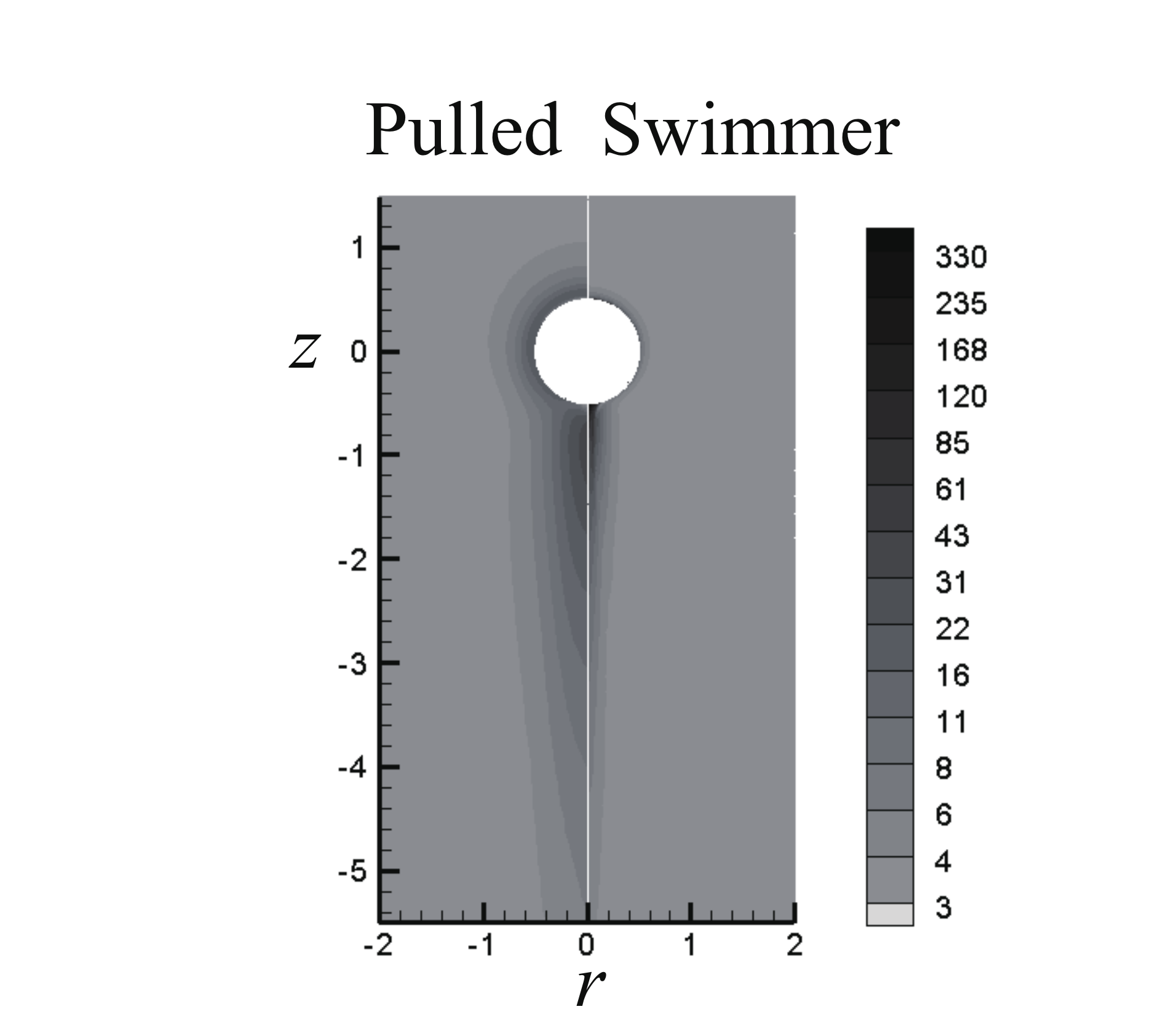}
   \caption{Polymeric stretching field: trace of the polymer conformation tensor, $Tr(\mathbf{C})$. 
   Comparison between forced motion (left) and free-swimming at the same speed (right), with $We=7$ and $\beta=0.3$. }
   \label{fig:poly_stretch_a}
\end{figure}

\begin{figure}[t]
   \centering
   \includegraphics[width=0.48 \textwidth]{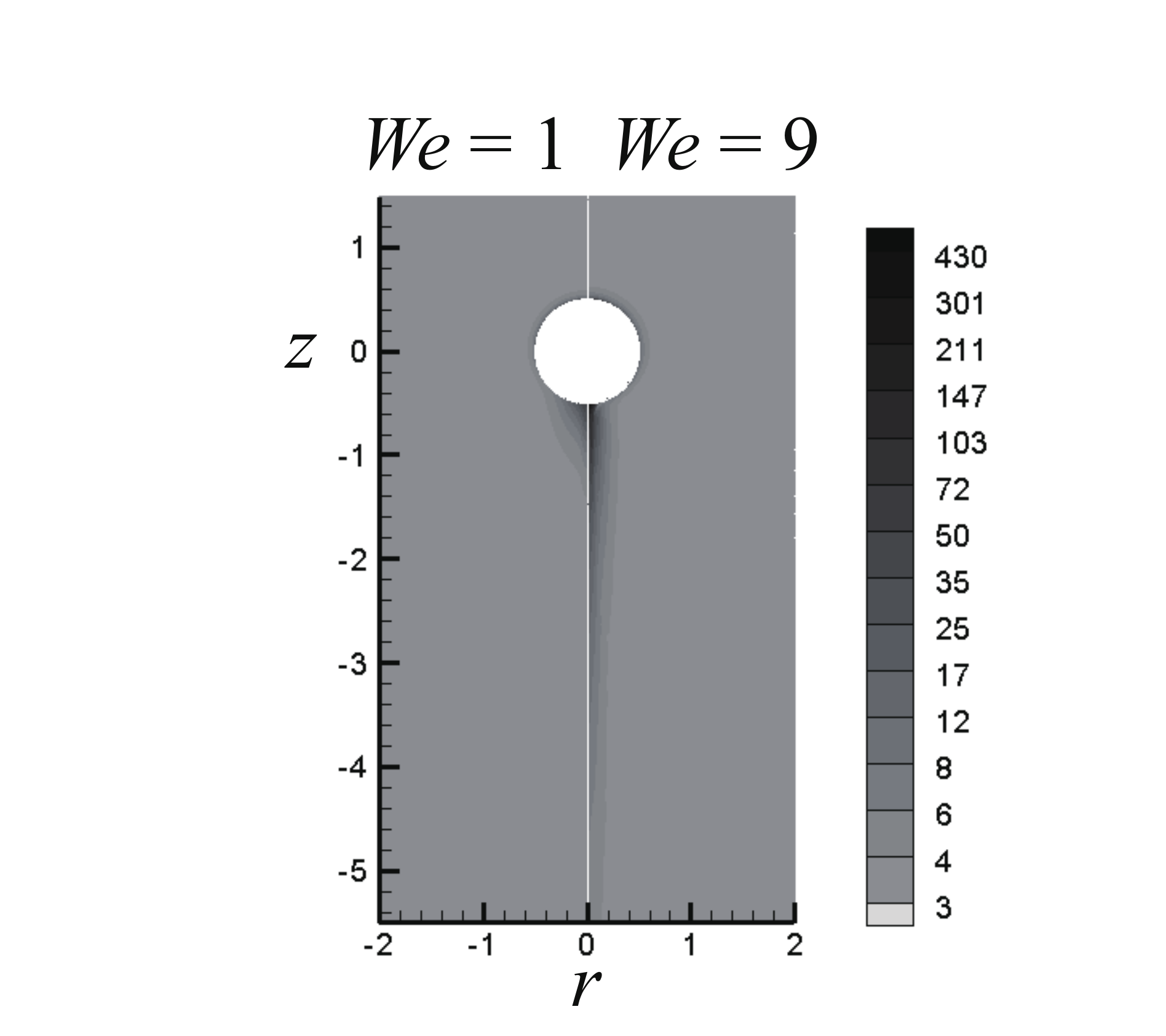}
   \caption{Polymeric stretching field: trace of the polymer conformation tensor, $Tr(\mathbf{C})$. 
   Comparison of polymer stretching between $We=1$ and $We=9$ for $\beta=0.3$. }
   \label{fig:poly_stretch_b}
\end{figure}

The trace of the polymer conformation tensor, $\mathbf{C}$, defined as
\begin{equation}
\mathbf{C}=\frac{We}{1-\beta} (\boldsymbol{\tau}^{p}+\mathbf{I}),
\end{equation}
 indicates the elongation of the polymers in the fluid. We plot $Tr (\mathbf{C})$ in Fig.~\ref{fig:poly_stretch_a} and \ref{fig:poly_stretch_b} for the forced motion of the sphere and for swimming with different polymer relaxation times. 
 In Fig.~\ref{fig:poly_stretch_a} we compare polymer stretching for forced motion and free swimming at the same speed in the case where  $We=7$ and $\beta=0.3$. 
 The region around the body where stretching is evident is much larger in the case of forced motion. The spatial decay of stretching is more rapid on the side of the swimmer while the largest elongation is observed in the wake right behind the organism. In Fig.~\ref{fig:poly_stretch_b} we  show the variation of stretching at  different values of $We$. As expected   a larger Weissenberg number leads to a larger region of  elongated polymers, and correlates with a more pronounced elastic wake.

\begin{figure}[t]
   \centering
   \includegraphics[width=0.48 \textwidth]{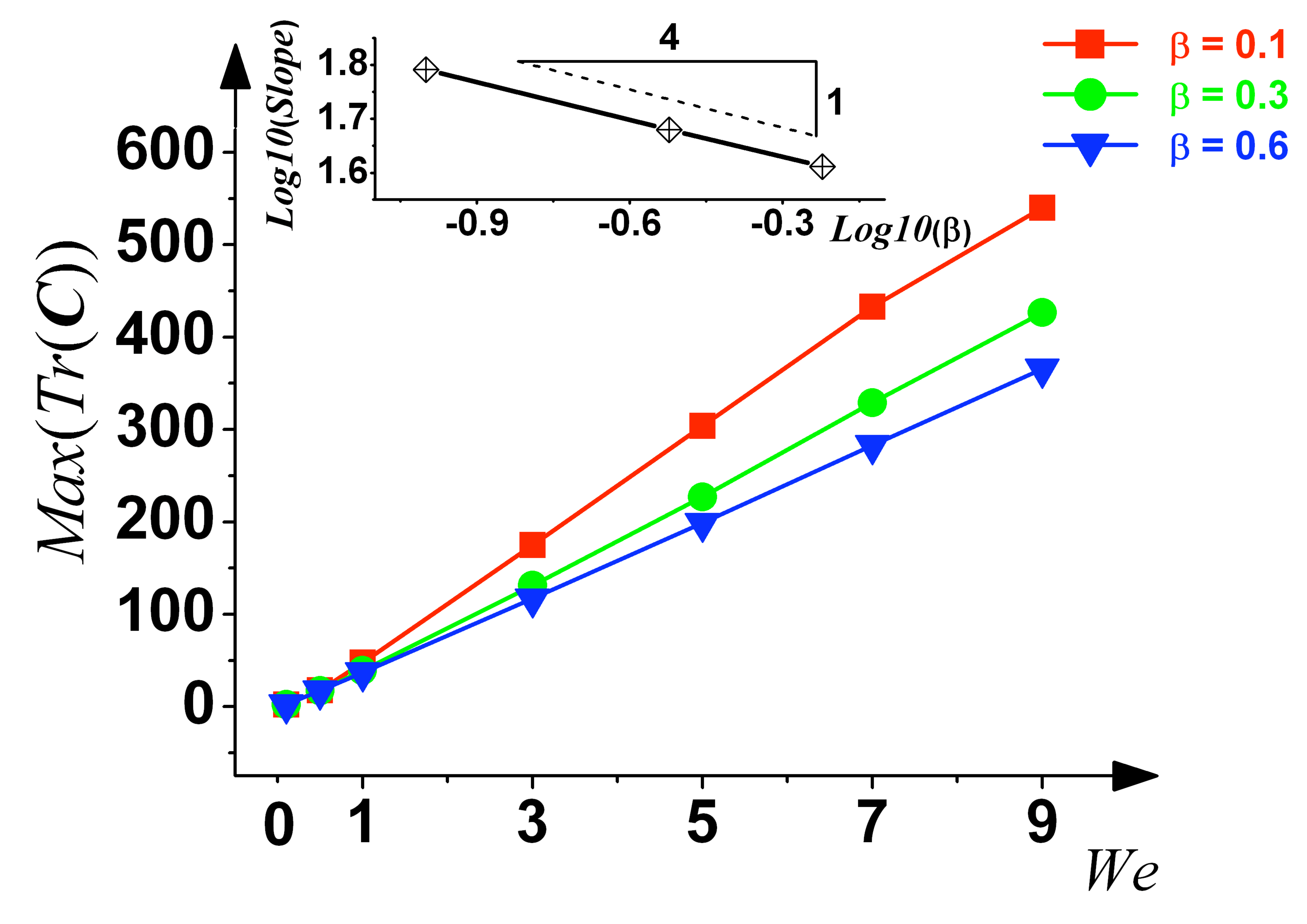}
   \caption{(Color online)  Dependence of the maximum polymer stretching, $Max(Tr(\mathbf{C}))$,  on the Weissenberg  number, $We$. In the range of Weissenberg  numbers considered, the relationship is approximately linear, with a slope $s$. Inset: dependance of the slope, $s$, on the viscosity ratio $\beta$ (log-log plot); the solid line is a guide to the eye showing a power law of 1/4.}
   \label{fig:poly_stretch_scaling}
\end{figure}

The increase of the magnitude of the polymer elongation is further quantified in Fig.~\ref{fig:poly_stretch_scaling} where the maximum of  $ Tr(\mathbf{C})$ inside our computational domain is displayed as a function of $We$ for the different values of $\beta$ considered. The relationship between elongation and relaxation time is found to be approximately linear, with a slope $s$ dependent on the viscosity ratio. The dependence of the slope with $\beta$ is shown in the inset in Fig.~\ref{fig:poly_stretch_scaling}, and a power law $s \approx \beta^{-1/4}$ provides an appropriate  fit to our numerical results.

\begin{table}[t]
\centering
\begin{tabular} {llcc c  cc cc  cc c  cc c cc c cc}
\hline
& & && \quad \quad $C_{rr}$\quad \quad & \quad \quad $C_{zz}$\quad \quad & \quad \quad $C_{rz}$ \quad \quad& \quad \quad $C_{\phi \phi}$\quad \quad \\
\hline
\\
$We=1$ \quad\quad && & & \bf  16.2 & \bf 39.3  & \bf -9.4  & \bf 16.3 \\
&$R$ && & 0.5  & 0.526  & 0.5     & 0.5 \\
&$\theta$  ($^\circ$)&&& 3.5 & 177  & 166  & 5 \\
\\
\hline
\\
$We=5$&  &&& \bf 83.8 & \bf 223.6 & \bf-38.3  & \bf 84.2 \\
&$R$ && &0.5  & 0.529  & 0.5     & 0.5 \\
&$\theta$  ($^\circ$)&& & 4 & 178  & 171  & 7 \\
\\
\hline
\\
$ We=9$  &&&& \bf 141.27 & \bf 426.6  & \bf -89.7 \bf & \bf 141.7 \\
&$R$ & &&0.5  & 0.528  & 0.5     & 0.5 \\
&$\theta$ ($^\circ$) && & 4 & 178.3  & 171.7  & 7.5 \\
\\
\hline
\end{tabular}
\caption{Maxima of individual components of polymer stretching (bold numbers) and corresponding location where these maxima are attained for different values of $We$, and in the case $\beta=0.3$. The position is reported in spherical polar coordinate, with $\theta$ in degrees and $R$ nondimensionalized by the sphere diameter, while the polymeric stresses are in cylindrical coordinate (see Fig.~\ref{fig:coordisys}).} 
\label{tab:elong}
\end{table}

Finally, we report in Table~\ref{tab:elong} the maxima of the different components of the conformation tensor ($C_{rr}$, $C_{zz}$, $C_{rz}$, $C_{\phi\phi}$) together with the location where these maxima are  attained (spherical coordinates with $R$ the distance from the center of the swimmer  and $\theta$ in degrees measured from the front of the swimmer).  Three values of the  Weissenberg number are considered  with $\beta=0.3$.  The maximum elongation is in  axial stretching, $C_{zz}$, and occurs just behind the body (see also Fig.~\ref{fig:poly_stretch_b}). The maximum of radial stretching $C_{rr}$ is observed on the swimmer, just off the symmetry line at the front stagnation, while the peak of the shear $C_{rz}$ is characterized by negative values and is observed on the back of the body with values of $\theta$ slightly increasing with polymer elasticity; this component will be responsible for the negative wake further downstream \cite{harlen02}.  In addition, and as expected, the  component $C_{\phi \phi}$  is also nonzero.  Its amplitude is in fact comparable to that of the radial stretching $C_{rr}$ and it attains its maximum value in front of the cylinder.

\section{Prolate swimmers}
\label{prolate}

After considering spherical bodies, we extend in this section our results 
to the case of prolate swimmers of different aspect ratios. We assume the body to be an axisymmetric prolate spheroid with an aspect ratio, $\mathcal{AR} > 1$, defined as the  ratio between its major (symmetry) axis, and its minor axis. In order to present a  proper comparison between organisms of different shapes,  we  keep their volume fixed. As a consequence, we adopt as reference length for our dimensionless numbers $2 \tilde{R}$, with $\tilde{R}={(3V/4\pi)}^{1/3}$;  $\tilde{R}$ is thus the radius of a sphere having same volume $V$ as the prolate ellipsoid. As an example, for a swimmer with aspect ratio $\mathcal{AR}=4$, the semi-major axis is  $0.315 D$, the semi-minor axis is $1.26 D$, and simulations are performed for the same values of $We$ and $\beta$ as for the sphere of diameter $D$.

\begin{figure}[t]
   \centering
   \includegraphics[width=0.5 \textwidth]{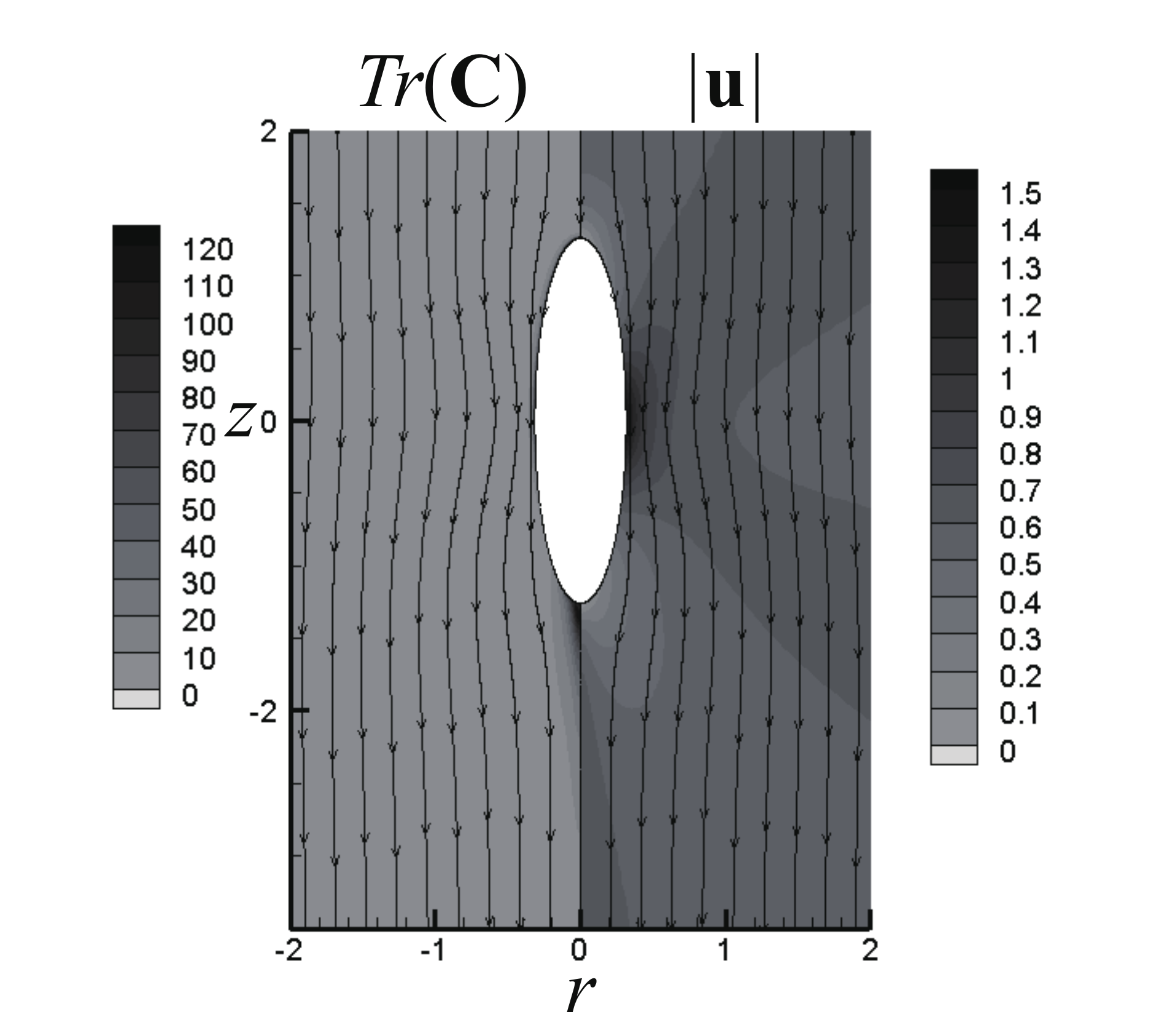}
   \caption{ Locomotion of a prolate swimmer of aspect ratio $\mathcal{AR}=4$ in a viscoelastic fluid with $We=7$ and $\beta=0.3$.
    Left: Trace of the conformation tensor, $Tr(\mathbf{C})$. Right: Velocity magnitude, $|\bf u|$. Streamlines are reported on both sides.}
   \label{fig:prolate}
\end{figure}

Computational results for swimming speed, power and efficiency as a function of the Weissenberg number and the viscosity ratio show similar trends as those discussed earlier for  spherical squirming, and will not be repetead. 
As example of flow, we show in Fig.~\ref{fig:prolate} the flow streamlines and polymer elongation for a prolate swimmer of aspect ratio $\mathcal{AR}=4$.  Large values of $Tr(\mathbf{C})$ are observed in a thin region around the body and in the wake, similarly to the spherical swimmer. The thickness of this stretching boundary layer, as well as the length of the wake, is found to decrease for an elongated swimmer.

Comparing the polymer stretching reported in Fig.~\ref{fig:poly_stretch_a} and in Fig.~\ref{fig:prolate}, we note also that the maximum of $Tr(\mathbf{C})$ is more than twice as big in the case of a spherical swimmer. In addition, for the prolate swimmer the velocity displays a weak overshoot just behind the body and, more interestingly, the streamlines are seen to converge toward the centre of the body ($z=0$),  and then depart further downstream. This is explained by the formation of a region of negative (resp. positive) pressure near the surface on the front (resp. rear) of the body. Such an antisymmetric pressure distribution with respect to the plan $z=0$ is not observed for spherical squirmers.

\begin{figure}[t]
   \centering
   \includegraphics[width=0.48 \textwidth]{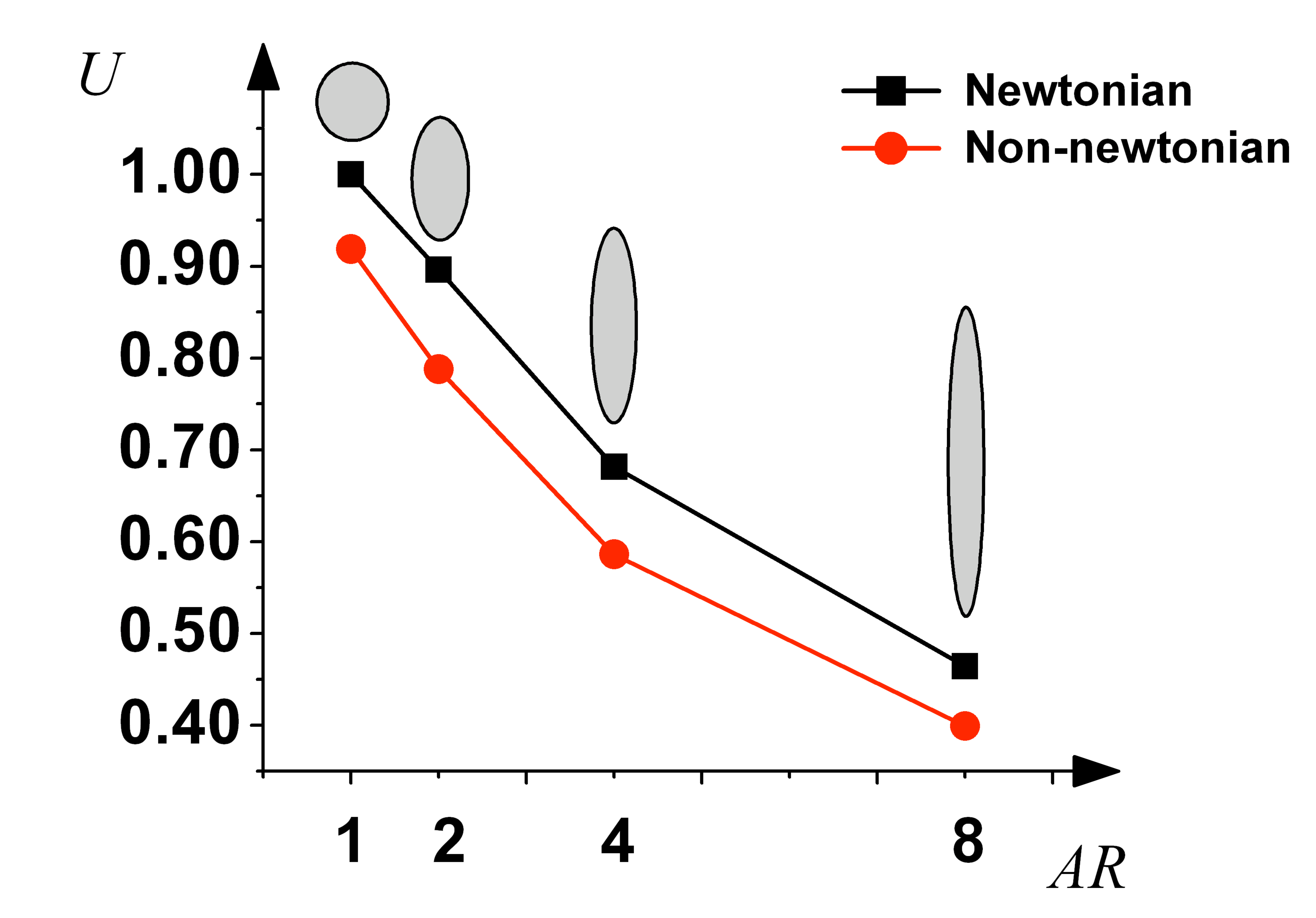}
   \caption{(Color online)  Swimming speed in the polymeric fluid with $We=7$ and $\beta=0.3$ divided by that of the spherical Newtonian swimmer for the prolate microorganism sharing the same volume but with different aspect ratio $\mathcal{AR}$.}
   \label{fig:prolate_speed}
\end{figure}

\begin{figure}[t]
   \centering
   \includegraphics[width=0.48 \textwidth]{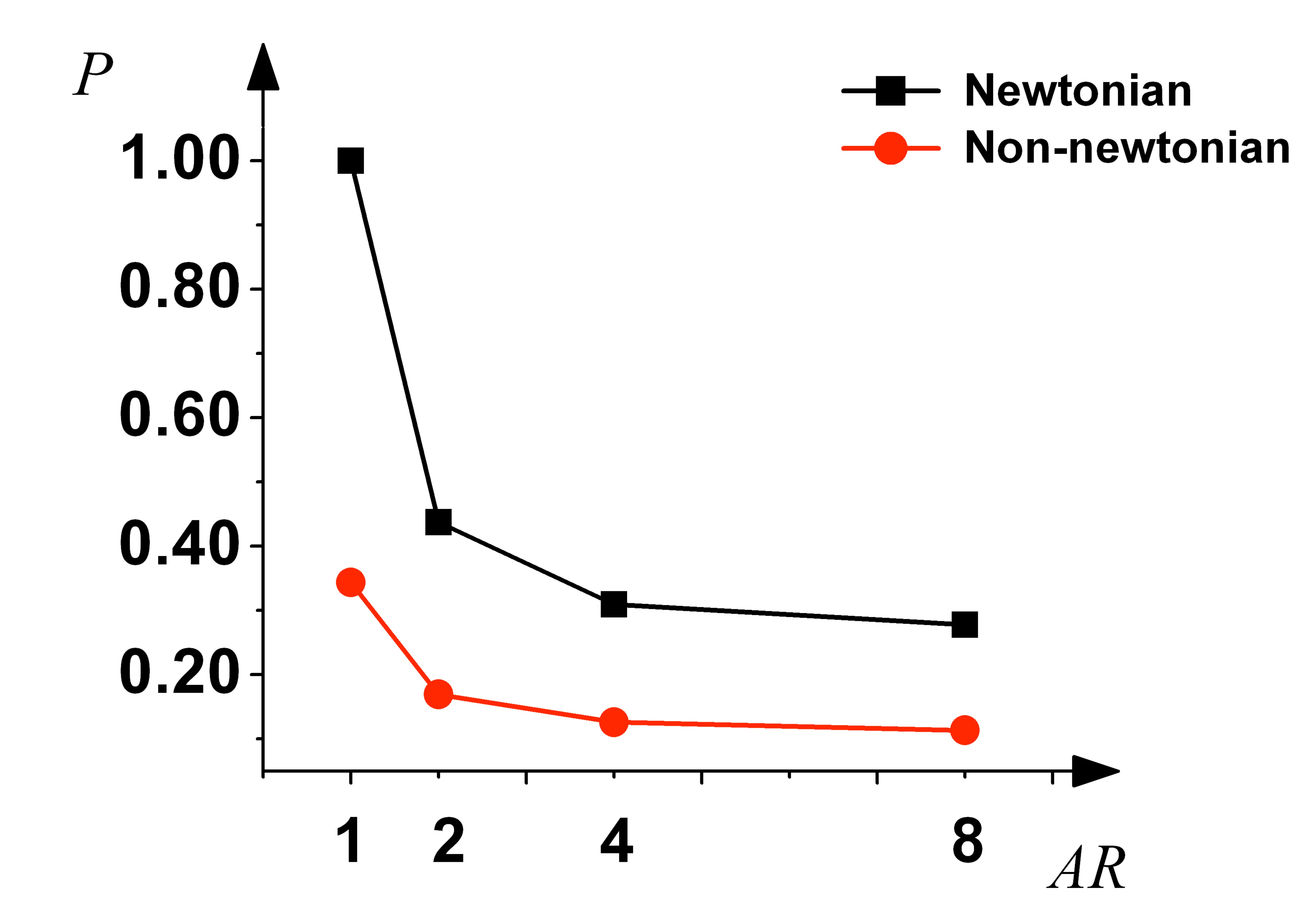}
   \caption{(Color online) Swimming power in the polymeric fluid with  $We=7$ and $\beta=0.3$ divided by that of the spherical Newtonian swimmer for the prolate microorganism sharing the same volume but with different aspect ratio $\mathcal{AR}$.}
   \label{fig:prolate_power}
\end{figure}

In Fig.~\ref{fig:prolate_speed} we show the variation of the swimming speed with the prolate aspect ratio. We plot the results in the Newtonian case  (black squares) as well as the polymeric case with  $We=7$ and $\beta=0.3$ (red circles).  The swimming speed is normalized with the swimming velocity of the spherical Newtonian squirmer, and is seen to  decrease with the aspect ratio. 
{To explain this finding we consider the pressure distribution around the organism. 
For the spherical squirmer, the two regions of minimum and maximum pressure are close to each other and on the rear of the body, whereas 
for the elongated squirmer we find high pressure on the front and low pressure on
the rear. This implies that the pressure forces act in the direction opposite to that of swimming for prolate organisms.}

The swimming power, normalized by that of the sphere in the Newtonian fluid with the same total viscosity, is shown in Fig.~\ref{fig:prolate_power}, and also decreases with the aspect ratio of the body.  The relative reduction in consumed power is increasing with decreasing aspect ratio.

\begin{figure}[b]
   \centering
   \includegraphics[width=0.48 \textwidth]{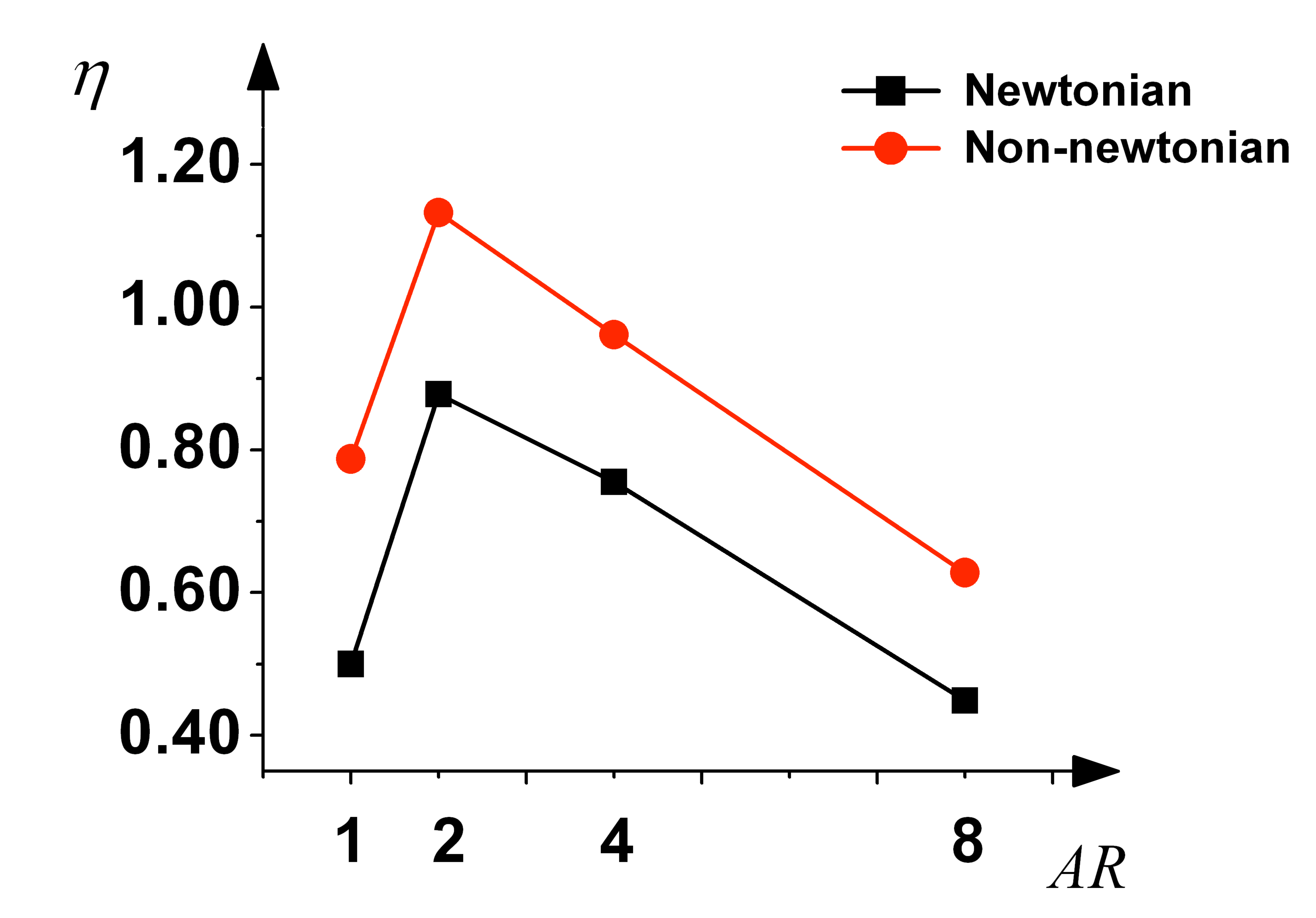}
   \caption{(Color online) Swimming efficiency of the prolate squirmer in a polymeric fluid as a function of aspect ratio $\mathcal{AR}$,  with $We=7$ and $\beta=0.3$.  
   Squirmers of different aspect ratio have the same volume.}
   \label{fig:prolate_eff}
\end{figure}

The swimming  efficiency is displayed in Fig.~\ref{fig:prolate_eff}. We find that the swimmer of aspect ratio  $\mathcal{AR}\approx 2$ is the most efficient, a result which is valid both in the Newtonian and non-Newtonian limit. 
In addition, a robust increase in efficiency in the viscoelastic fluid is also evident.

\section{Conclusion}
\label{discussion}

Although significant progress has been made in the analysis of low-Reynolds number locomotion in Newtonian fluids, many biological cells   encounter viscous environments with suspended microstructures or macromolecules. It  is thus of fundamental importance to develop modeling tools addressing the effect of non-Newtonian stresses on propulsion.  In our study, we presented the results of numerical simulations  for a steady squirmer free-swimming in  a model (Giesekus) polymeric fluid. Locomotion is achieved by a prescribed steady tangential surface deformation of the body, which thus displays no shape change. To the best of our knowledge, the results discussed in our paper  are the first three-dimensional simulations for self-propelled motion in a complex fluid. In addition, as stresses in the Giesekus model saturate for large elongation, our results  are relevant to cell locomotion in concentrated polymeric solutions,  and long polymer relaxation times, as demonstrated by the appearance of negative wakes behind the swimmer.

Our main results are as follows. We first showed that the swimming speed is lower than in a Newtonian fluid, with a minimum near $We\approx 1$. Swimming power also decreases with polymer relaxation time and increased viscosity contrast between the polymer and the solvent. 
{ Rescaling the data, it is possible to show when keeping constant the consumed power, the velocity at the boundary increases and the swimming speed increases. Swimming at constant power gives therefore larger speeds in viscoelastic fluids.}  The swimming efficiency, defined as the ratio between the power required to pull the swimmer in the same fluid at the same speed, and the power consumed by the swimmer, is 
 found to systematically increase in viscoelastic fluids. The gain in efficiency is larger for the longest relaxation times and higher polymer viscosity contrast, and  approaches a constant asymptotic value at high Weissenberg number. The increase in efficiency is consistent with the analysis in Ref.~\cite{leshansky09} for a squirmer in a suspension of rigid spheres and with the numerical results of Ref.~\cite{teran_PRL} for a two-dimensional sheet of finite length. Flow visualizations further reveal that the fore-aft symmetry of a Newtonian swimmer is broken in a viscoelastic fluid. The appearance of a negative elastic wake, i.e. one with velocity directed towards the object, is numerically demonstrated. The analysis of the main flow features indicates that the swimming speed of a viscoelastic squirmer increases for $We\gtrsim 1$ when  
the elastic wake moves further downstream, and the polymer molecules experience larger elongations. 
 The minimum speed is observed at $We\approx 1$ when the distance travelled by the polymer during their relaxation time is of the order of the body length. In this case, a region of strong normal stress $\tau_{zz}$ is forming on the back of the swimmer, pulling it backward. This region is weaker for lower $We$ and located further downstream for larger $We$.
 The spatial decay rate of the flow induced by the swimmer is found to be larger  in a viscoelastic flow,  suggesting different collective behavior of active suspensions of self-propelled bodies  in Newtonian vs. non-Newtonian fluids.  Finally, we extended our results to the case of prolate swimmers of different aspect ratios, and found that bodies with an aspect ratio of two have the  largest swimming efficiency. The gain in efficiency provided by the fluid viscoelasticity is more pronounced for the spherical swimmer. This suggests that bluff bodies have larger possibilities to exploits the advantages of such an environment.

The work presented in this paper could be extended in a number of non-trivial and interesting ways. First, we have focused here on a single squirming mode, but most swimming cells have a dipolar nature in the far field, and thus the case $\beta_{SW}\neq 0$ should be investigated next.  In particular, the distinction between pushers and pullers should be addressed. Second, real cells do not deform the surrounding fluid in a steady fashion, but usually apply time-dependent kinematics, and these unsteady effects, when they occur for Deborah numbers of order unity or above, can in general not be neglected. Third, more realistic cell geometries should be considered, in particular for flagellated bacteria. Fourth, the breakdown of the front-back flow symmetry and the appearance of wakes will surely have interesting consequences on hydrodynamic interactions between swimmers. 
Finally, most cells do not swim with prescribed kinematics, but instead their deformation kinematics are obtained as a physical balance between an internal (or boundary) actuation and the outside fluid, and this coupled fluid-solid problem should be further investigated.

\section*{Acknowledgments}
Funding  by VR (the Swedish Research Council) and the National Science Foundation (grant  CBET-0746285 to E.L.) is gratefully acknowledged.
Computer time provided by SNIC (Swedish National Infrastructure for Computing) is also acknowledged.

\bibliographystyle{unsrt}
\bibliography{bionon,luca}

\end{document}